\documentclass[usenatbib]{emulateapj}
\pdfoutput=1
\shorttitle{The Dark Matter distribution in a cluster at $z=1$}
\shortauthors{Collett et al. (2016)}

\usepackage{color}			      % using color package        - for emulateapj
\definecolor{midgray}{gray}{0.4}		% defining gray color        - for emulateapj
\definecolor{orange}{rgb}{1,0.5,0}    %        - for emulateapj

\usepackage{xspace}
\usepackage{graphicx}
\usepackage{amsmath}
\newcommand{\new}[1]{ { {#1}} }
\def\spose#1{\hbox  to 0pt{#1\hss}}  
\newcommand{\lta}{\mathrel{\spose{\lower 3pt\hbox{$\sim$}}\raise  2.0pt\hbox{$<$}}}
\newcommand{\gta}{\mathrel{\spose{\lower  3pt\hbox{$\sim$}}\raise 2.0pt\hbox{$>$}}}

\newcommand{\be}{\begin{equation}}
\newcommand{\ee}{\end{equation}}

\newcommand{\citepeg}[1]{\citep[e.g.][]{#1}}

\newcommand{\kms}{\ifmmode  \,\mathrm {km}\,s^{-1} \else $\,\mathrm{ km\,s^{-1} } $ \fi }
\newcommand{\kpc}{\ifmmode  {\mathrm{kpc}}  \else ${\mathrm{  kpc}}$ \fi  }  
\newcommand{\pc}{\ifmmode  {\mathrm{ pc}}  \else ${\mathrm{ pc}}$ \fi  }  
\newcommand{\Msun}{\ifmmode {\mathrm{ M_{\odot}}} \else ${\mathrm{ M_{\odot}}}$ \fi} 
\newcommand{\Zsun}{\ifmmode {\mathrm{{Z_{\odot}}}} \else ${\mathrm{ Z_{\odot}}}$ \fi} 
\newcommand{\yr}{\ifmmode yr^{-1} \else $yr^{-1}$ \fi} 
\newcommand{\hMsun}{\ifmmode h^{-1}\,\rm M_{\odot} \else $h^{-1}\,mathrm{\ M_{\odot}}$ \fi}

\def\spose#1{\hbox  to 0pt{#1\hss}}  
\renewcommand{\lta}{\mathrel{\spose{\lower 3pt\hbox{$\sim$}}\raise  2.0pt\hbox{$<$}}}
\renewcommand{\gta}{\mathrel{\spose{\lower  3pt\hbox{$\sim$}}\raise 2.0pt\hbox{$>$}}}

\renewcommand{\be}{\begin{equation}}
\renewcommand{\ee}{\end{equation}}

\newcommand{\bea}{\begin{eqnarray}}
\newcommand{\eea}{\end{eqnarray}}

\newcommand{\comment}[1]{}
\newcommand{\comments}[1]{}
\newcommand{\angstrom}{\mbox{\normalfont\AA}}

\begin{document}

\title{Core or cusps: The central dark matter profile of a redshift one strong lensing cluster with a bright central image}

\author{Thomas~E.~Collett$^{1}$,
Elizabeth~Buckley-Geer$^{2}$,
Huan~Lin$^{2}$,
David~Bacon$^{1}$,
Robert~C.~Nichol$^{1}$,
Brian~Nord$^{2}$,
Xan~Morice-Atkinson$^{1}$,
Adam~Amara$^{3}$,
Simon~Birrer$^{4,3}$,
Nikolay~Kuropatkin$^{2}$,
Anupreeta~More$^{5}$,
Casey~Papovich$^{6}$,
Kathy~K.~Romer$^{7}$,
Nicolas~Tessore$^{8}$,
Tim M. C.~Abbott$^{9}$,
Sahar~Allam$^{2}$,
James~Annis$^{2}$,
Aurélien~Benoit-L{\'e}vy$^{10,11,12}$,
David~Brooks$^{11}$,
David~L.~Burke$^{13,14}$,
Matias~Carrasco~Kind$^{15,16}$,
Francisco Javier~J.~Castander$^{17}$,
Chris~B.~D'Andrea$^{18}$,
Luiz~N.~da Costa$^{19,20}$,
Shantanu~Desai$^{21}$,
H. Thomas~Diehl$^{2}$,
Peter~Doel$^{11}$,
Tim~F.~Eifler$^{22,23}$,
Brenna~Flaugher$^{2}$,
Josh~Frieman$^{2,24}$,
David~W.~Gerdes$^{25,26}$,
Daniel~A.~Goldstein$^{27,28}$,
Daniel~Gruen$^{13,14}$,
Julia~Gschwend$^{19,20}$,
Gaston~Gutierrez$^{2}$,
David~J.~James$^{29,9}$,
Kyler~Kuehn$^{30}$,
Steve~Kuhlmann$^{31}$,
Ofer~Lahav$^{11}$,
Ting~S.~Li$^{2,6}$,
Marcos~Lima$^{32,19}$,
Marcio~A.~G.~Maia$^{19,20}$,
Marisa~March$^{18}$,
Jennifer~L.~Marshall$^{6}$,
Paul~Martini$^{33,34}$,
Peter~Melchior$^{35}$,
Ramon~Miquel$^{36,37}$,
Andrés~A.~Plazas$^{23}$,
Eli~S.~Rykoff$^{13,14}$,
Eusebio~Sanchez$^{38}$,
Vic~Scarpine$^{2}$,
Rafe~Schindler$^{14}$,
Michael~Schubnell$^{26}$,
Ignacio~Sevilla-Noarbe$^{38}$,
Mathew~Smith$^{39}$,
Flavia~Sobreira$^{40,19}$,
Eric~Suchyta$^{41}$,
Molly~E.~C.~Swanson$^{16}$,
Gregory~Tarle$^{26}$,
Douglas~L.~Tucker$^{2}$,
Alistair~R.~Walker$^{9}$
}\affil{$^{1}$Institute of Cosmology \& Gravitation, University of Portsmouth, Portsmouth, PO1 3FX, UK\\$^{2}$Fermi National Accelerator Laboratory, P. O. Box 500, Batavia, IL 60510, USA\\$^{3}$Department of Physics, ETH Zurich, Wolfgang-Pauli-Strasse 16, CH-8093 Zurich, Switzerland\\$^{4}$Department of Physics and Astronomy, UCLA, PAB, 430 Portola Plaza, Box 951547, Los Angeles, CA 90095-1547, USA\\$^{5}$Kavli IPMU (WPI), UTIAS, The University of Tokyo, Kashiwa, Chiba 277-8583, Japan\\$^{6}$George P. and Cynthia Woods Mitchell Institute for Fundamental Physics and Astronomy, and Department of Physics and Astronomy, Texas A\&M University, College Station, TX 77843,  USA\\$^{7}$Department of Physics and Astronomy, Pevensey Building, University of Sussex, Brighton, BN1 9QH, UK\\$^{8}$Jodrell Bank Center for Astrophysics, School of Physics and Astronomy, University of Manchester, Oxford Road, Manchester, M13 9PL, UK\\$^{9}$Cerro Tololo Inter-American Observatory, National Optical Astronomy Observatory, Casilla 603, La Serena, Chile\\$^{10}$CNRS, UMR 7095, Institut d'Astrophysique de Paris, F-75014, Paris, France\\$^{11}$Department of Physics \& Astronomy, University College London, Gower Street, London, WC1E 6BT, UK\\$^{12}$Sorbonne Universit\'es, UPMC Univ Paris 06, UMR 7095, Institut d'Astrophysique de Paris, F-75014, Paris, France\\$^{13}$Kavli Institute for Particle Astrophysics \& Cosmology, P. O. Box 2450, Stanford University, Stanford, CA 94305, USA\\$^{14}$SLAC National Accelerator Laboratory, Menlo Park, CA 94025, USA\\$^{15}$Department of Astronomy, University of Illinois, 1002 W. Green Street, Urbana, IL 61801, USA\\$^{16}$National Center for Supercomputing Applications, 1205 West Clark St., Urbana, IL 61801, USA\\$^{17}$Institut de Ci\`encies de l'Espai, IEEC-CSIC, Campus UAB, Carrer de Can Magrans, s/n,  08193 Bellaterra, Barcelona, Spain\\$^{18}$Department of Physics and Astronomy, University of Pennsylvania, Philadelphia, PA 19104, USA\\$^{19}$Laborat\'orio Interinstitucional de e-Astronomia - LIneA, Rua Gal. Jos\'e Cristino 77, Rio de Janeiro, RJ - 20921-400, Brazil\\$^{20}$Observat\'orio Nacional, Rua Gal. Jos\'e Cristino 77, Rio de Janeiro, RJ - 20921-400, Brazil\\$^{21}$Department of Physics, IIT Hyderabad, Kandi, Telangana 502285, India\\$^{22}$Department of Physics, California Institute of Technology, Pasadena, CA 91125, USA\\$^{23}$Jet Propulsion Laboratory, California Institute of Technology, 4800 Oak Grove Dr., Pasadena, CA 91109, USA\\$^{24}$Kavli Institute for Cosmological Physics, University of Chicago, Chicago, IL 60637, USA\\$^{25}$Department of Astronomy, University of Michigan, Ann Arbor, MI 48109, USA\\$^{26}$Department of Physics, University of Michigan, Ann Arbor, MI 48109, USA\\$^{27}$Department of Astronomy, University of California, Berkeley,  501 Campbell Hall, Berkeley, CA 94720, USA\\$^{28}$Lawrence Berkeley National Laboratory, 1 Cyclotron Road, Berkeley, CA 94720, USA\\$^{29}$Astronomy Department, University of Washington, Box 351580, Seattle, WA 98195, USA\\$^{30}$Australian Astronomical Observatory, North Ryde, NSW 2113, Australia\\$^{31}$Argonne National Laboratory, 9700 South Cass Avenue, Lemont, IL 60439, USA\\$^{32}$Departamento de F\'{\i}sica Matem\'atica,  Instituto de F\'{\i}sica, Universidade de S\~ao Paulo,  CP 66318, CEP 05314-970, S\~ao Paulo, SP,  Brazil\\$^{33}$Center for Cosmology and Astro-Particle Physics, The Ohio State University, Columbus, OH 43210, USA\\$^{34}$Department of Astronomy, The Ohio State University, Columbus, OH 43210, USA\\$^{35}$Department of Astrophysical Sciences, Princeton University, Peyton Hall, Princeton, NJ 08544, USA\\$^{36}$Instituci\'o Catalana de Recerca i Estudis Avan\c{c}ats, E-08010 Barcelona, Spain\\$^{37}$Institut de F\'{\i}sica d'Altes Energies (IFAE), The Barcelona Institute of Science and Technology, Campus UAB, 08193 Bellaterra (Barcelona) Spain\\$^{38}$Centro de Investigaciones Energ\'eticas, Medioambientales y Tecnol\'ogicas (CIEMAT), Madrid, Spain\\$^{39}$School of Physics and Astronomy, University of Southampton,  Southampton, SO17 1BJ, UK\\$^{40}$Instituto de F\'isica Gleb Wataghin, Universidade Estadual de Campinas, 13083-859, Campinas, SP, Brazil\\$^{41}$Computer Science and Mathematics Division, Oak Ridge National Laboratory, Oak Ridge, TN 37831\\}

\email{thomas.collett@port.ac.uk}
%\author{Thomas E. Collett$^1$,  Elizabeth Buckley-Geer$^2$, Huan Lin$^2$, David Bacon$^1$, Robert C. Nichol$^1$, Brian Nord$^2$, Xan Morice-Atkinson$^1$, Adam Amara$^{3}$, Simon Birrer$^{3,4}$, Nikolay Kuropatkin$^{2}$, Anupreeta More$^{5}$, Casey Papovich$^{6,7}$, Kathy Romer, Nicolas Tessore}

%%%%%%%%%%%%%%%%%%%%%%%%%%%%%%%%%%%%%%%%%%%%%%%%%%%%%%%%%%%%%%%%%%%%%%%%

%\pagerange{\pageref{firstpage}--\pageref{lastpage}}\pubyear{2015}

%\maketitle 

%\label{firstpage}

\begin{abstract}

We report on SPT-CLJ2011-5228, a giant system of arcs created by a cluster at $z=1.06$. The arc system is notable for the presence of a bright central image. The source is a Lyman Break galaxy at $z_s=2.39$ and the mass enclosed within the 14 arc second radius Einstein ring is $\sim 10^{14.2}\Msun$. We perform a full light profile reconstruction of the lensed images to precisely infer the parameters of the mass distribution. The brightness of the central image demands that the central total density profile of the lens be shallow. By fitting the dark matter as a generalized Navarro-Frenk-White profile---with a free parameter for the inner density slope---we find that the break radius is $270^{+48}_{-76}$ kpc, and that the inner density falls with radius to the power $-0.38\pm0.04$ at 68 percent confidence. Such a shallow profile is in strong tension with our understanding of relaxed cold dark matter halos; dark matter only simulations predict the inner density should fall as $r^{-1}$. The tension can be alleviated if this cluster is in fact a merger; a two halo model can also reconstruct the data, with both clumps (density going as $r^{-0.8}$ and $r^{-1.0}$) much more consistent with predictions from dark matter only simulations. At the resolution of our Dark Energy Survey imaging, we are unable to choose between these two models, but we make predictions for forthcoming Hubble Space Telescope imaging that will decisively distinguish between them.

\end{abstract}


\setcounter{footnote}{1}

%%%%%%%%%%%%%%%%%%%%%%%%%%%%%%%%%%%%%%%%%%%%%%%%%%%%%%%%%%%%%%%%%%%%%%%%%%%%%%

\section{Introduction}
\label{sec:intro}

Cosmological cold dark matter (CDM) simulations suggest a model of hierarchical structure formation that produces a cuspy dark matter (DM) profile. \citet{NFW} found that the halo profiles are universal and well fit by the form
\be
\rho_{\mathrm{NFW}}(r) = \frac{\rho_0}{r (r_s+r)^2}.
\ee
The mass of a Navarro-Frenk-White (NFW) halo and the scale radius are the only free parameters, and CDM simulations show that these parameters are connected---with some scatter---by a mass--concentration relation \citep{duffy2010,neto2007}. However, baryonic physics complicates this simple cold dark matter picture \citep{gnedin2004}: gas can radiatively cool and contract, dragging DM inwards and increasing the central density, but supernovae, active galactic nuclei and dynamical heating can inject energy into the interstellar medium, causing the halo to expand \citep{pontzen+governato2012, teyssier2011, abadi2010,laporte2012}. Halos will also deviate from NFW if the dark matter is self interacting or if it has a large de Broglie wave length. In these cosmologies the dark matter self interaction creates a pressure that leads to the formation of central cores rather than high density cusps \citep{peter2013,rocha2013,Vogelsberger}. The central density profiles of halos are therefore sensitive probes of both baryonic feedback effects and the CDM paradigm.

Observationally, it has been shown that NFW profiles do not reproduce the central regions of halos over a range of masses. On the scales of dwarf galaxies there is evidence for central dark matter cores \citep{deblok2001,amorisco2013} derived from kinematic constraints, however recent simulations have cast doubt on  the interpretation of these data \citep{pineda2017}. On the scale of elliptical galaxies there is evidence for cuspy dark matter profiles that are much steeper than NFW. \citet{sonnenfeld2012} find $\rho\sim r^{-1.7}$ for an individual strong lensing galaxy with multiple background sources, and \citet{grillo2012} found that the average inner DM slope of 38 strong lenses is close to isothermal although these results depend on the assumed stellar mass-to-light ratio.

On the scale of galaxy clusters the central profiles are often seen to be flatter than NFW. \citet{sand} found that $\rho \sim r^{-0.5}$ for a sample of 6 strong lensing clusters. Work by \citet{newman2013a} combining stellar kinematics with strong and weak lensing found that the total mass profile in their clusters was consistent with NFW. However, because the total mass profile in those clusters is dominated by the baryonic content of the brightest cluster galaxy (BCG) they inferred that the dark matter profile was significantly shallower than NFW \citep{newman2013b}. However either a shallow central dark matter cusps or a cored NFW profile were able to reconstruct the data. Recently \citet{oldham+auger2016} showed that the dark matter halo of M87 has a 19 kpc core although the core size is degenerate with orbital anisotropy.

The primary difficulty of previous work on clusters has been the presence of a massive galaxy that dominates the total mass profile at the centre of the cluster. Inference on the dark halo are therefore only as robust as the subtraction of the stellar component. 

\begin{figure}
  \centering
    \includegraphics[width=\columnwidth,clip=True]{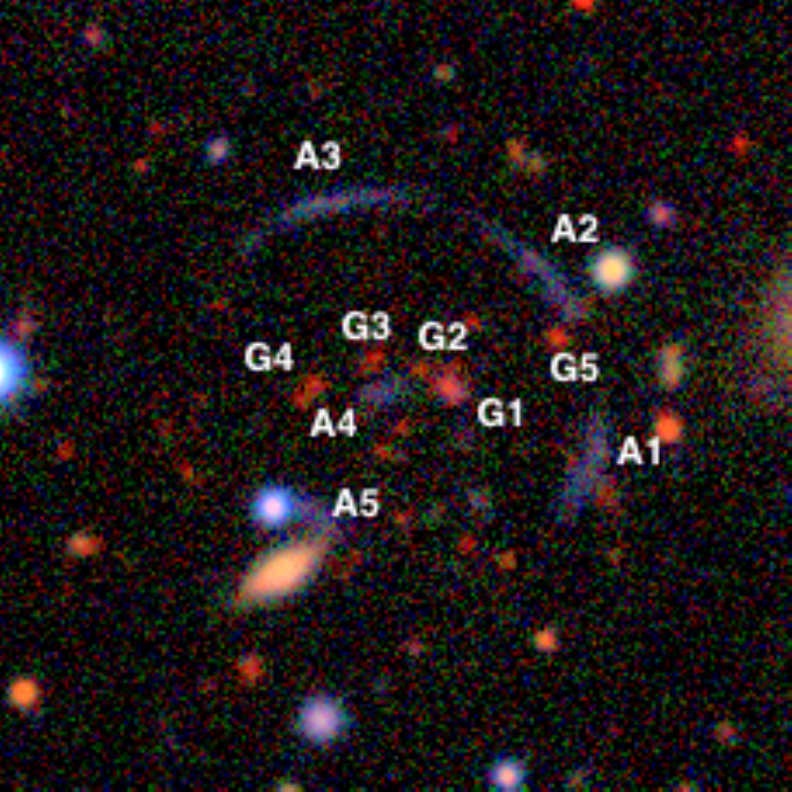}
    \caption {Pseudo-colour $gri$ composite image of the lens J2011, taken from the first three years of operation of DES. The image is 1 arcminute on a side.}
    \label{fig:colourimage}
\end{figure}

In this work we present imaging, spectroscopy and modelling of SPT-CLJ2011-5228 (hereafter J2011). J2011 is a $M_{500}=2.25 \pm 0.89 \times10^{14} h^{-1}_{70}\Msun$ cluster \citep{reichardt13} at $z=1.064$ acting as a strong gravitational lens with an  Einstein radius of $14.01 \pm 0.06$ arcsecond (Figure \ref{fig:colourimage}). The arc system is notable due to the presence of a bright central image. Central images are typically highly demagnified since in dense regions a small perturbation to the impact parameter of a light ray will produce a large change to the deflection angle. To date very few true central images are known \citep{sharon,inada2005,winn2004,colley1996}, and are typically highly demagnified. The presence of a bright central image in J2011 is therefore an exciting discovery and opens up the possibility to precisely constrain the central dark matter profile of this cluster.

In Section \ref{sec:data} of this paper we present our observations of J2011 using the Dark Energy Camera on the 4m Victor M. Blanco telescope and the GMOS spectrograph on the 8m Gemini South telescope. In Section \ref{sec:modelling}, we describe our lens modelling approach for this lens and present the result of two models for the dark matter distribution in the lens; a one halo model representing a relaxed cluster scenario in Section \ref{sec:1halo} and a two halo model representing a merging scenario in \ref{sec:2halo}. We provide interpretation of the results in Section \ref{sec:interpretation} and conclude in Section \ref{sec:conclusion}. \new{Unless otherwise stated, coordinates are measured West, North relative to 20:11:10.611, -52:28:40.12 (J2000) with a pixel size of 0.263 arcseconds.} Throughout this work we assume a flat $\Lambda$CDM cosmology with $H_0=67.8$ kms$^{-1}$Mpc$^{-1}$ and $\Omega_{\mathrm{M}}=0.308$ \citep{planckcosmo}.

%%%%%%%%%%%%%%%%%%%%%%%%%%%%%%%%%%%%%%%%%%%%%%%%%%%%%%%%%%%%%%%%%%%%%%%%

\section{Discovery of a giant arc system in DES and SPT data}
\label{sec:data}
J2011 was first reported as one of the 224 galaxy clusters detected in the first 720 ${\rm deg}^{2}$ of the South Pole Telescope (SPT) Sunyaev-Zel'dovich (SZ) survey. The SPT data selection is described in \citet{reichardt13}.  The cluster has a reported mass of $M_{500}=2.59 \pm 0.73 \times10^{14} h^{-1}_{70}\Msun$ with a detection significance of 4.58 \citep{bleem2015}. The SPT collaboration carried out follow-up optical and NIR imaging of these clusters using various telescopes and instruments as described in \citet{song12}. In Table 3 of \citet{song12} this cluster is noted as having a strong lensing arc. This area of the SPT footprint was imaged as part of the Year 1 (Y1) observations \citep{des,flaugher2015} of the Dark Energy Survey \citep[DES]{flaugher2005}. The DES subsequently rediscovered the system in a targeted visual search of all known SPT clusters in the Y1 footprint and separately in a visual search of red sources with multiple neighbours (H. T. Diehl in prep). The visual scans were performed on false-color PNG images, which are made by combining $g,r,i$ coaddition tiles into color images \citep{dacosta}.

\subsection{DES Imaging}

\begin{figure}
  \centering
    \includegraphics[width=\columnwidth,clip=True,trim={5mm 5mm 5mm 5mm}]{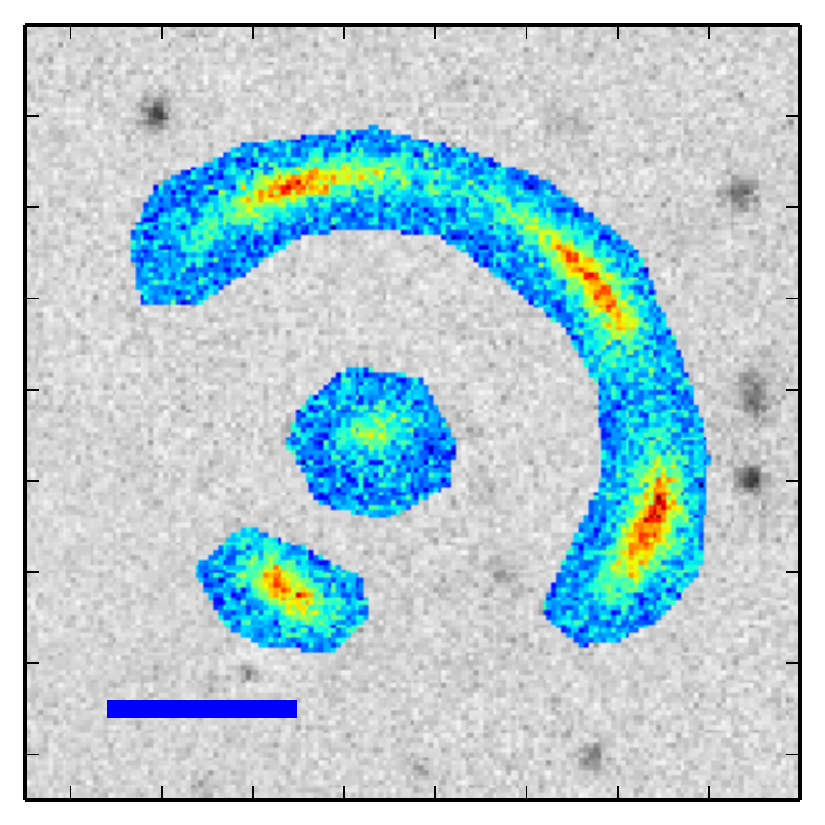}

    \caption {$g$ band image of the arcs and central image, after subtracting foregrounds. Only the coloured pixels are included in the lens modelling of Section \ref{sec:modelling}. The blue bar shows a ten arcsecond scale.}
    \label{fig:gsub}
\end{figure}

The lens modelling requires as input an image of the lensed arcs, a sigma image and a point-spread function (PSF) for each band. We use imaging from the first three years of DES to provide this data. The DESDM image processing pipeline (summarized in \citealp{balbinot2015} and described in detail in \citealp{sevilla2011, desai2012, mohr2012}) carries out image coaddition for each band and creates a sigma image which is the total uncertainty on each pixel of the coadd image due to noise. The DESDM pipeline also fits a model of the PSF using PSFEX  \citep{bertin2011,bertin1996}. The PSF model is used to generate a model star at the location of the cluster. The default DESDM sigma image only includes the noise from the sky photons and not the photons contributing to the objects in the image. We therefore modified and re-ran the image processing pipeline to generate a sigma image that also includes shot noise from the objects.

With our data, the lensed arc is brightest in the $g$-band, where there is negligible contribution to the flux from the cluster members, since the Balmer break lies redward of the $g$-band filter. There is however g-band emission close to arcs associated with two foreground stars (close to A2 and A5), a foreground galaxy (close to A5) plus emission from the BCG (which is close to the central image A4) (potentially produced by an AGN). These foregrounds are fitted with point sources for the stars, and S\'ersic profiles for the BCG and foreground galaxy. To ensure flux from the arcs is not attributed to the foregrounds, we simultaneously model the foregrounds and arcs using \new{an elliptical powerlaw density profile for the lens} \citep{barkana} and a parametric source with three Sersic components\footnote{The powerlaw lens model fits the outer arcs well but does not reproduce the central image.}. This lens model is then optimized and the g-band foregrounds subtracted. We use the foreground subtracted $g$-band data for our lens modelling. The foreground subtracted $g$-band data is shown in Figure \ref{fig:gsub}; it is this data that we use for the lens modelling in Section \ref{sec:modelling}.

\subsection{Gemini Spectroscopy of J2011}
We obtained spectroscopic follow-up observations with the Gemini Multi-Object Spectrograph \citep[GMOS]{gmos} on the Gemini South Telescope, as part of the Gemini Large and Long Program GS-2014B-LP-5\footnote{\url{http://www.gemini.edu/?q=node/12238\#Buckley}}. 

\begin{figure}
  \centering
    %\plottwo{mask_arcs.pdf}{mask_cluster.pdf}
    %\includegraphics[width=0.6\columnwidth,clip=True]{mask_arcs.pdf}\\
    %\includegraphics[width=0.6\columnwidth,clip=True]{mask_cluster.pdf}
    \includegraphics[width=0.6\columnwidth,clip=True]{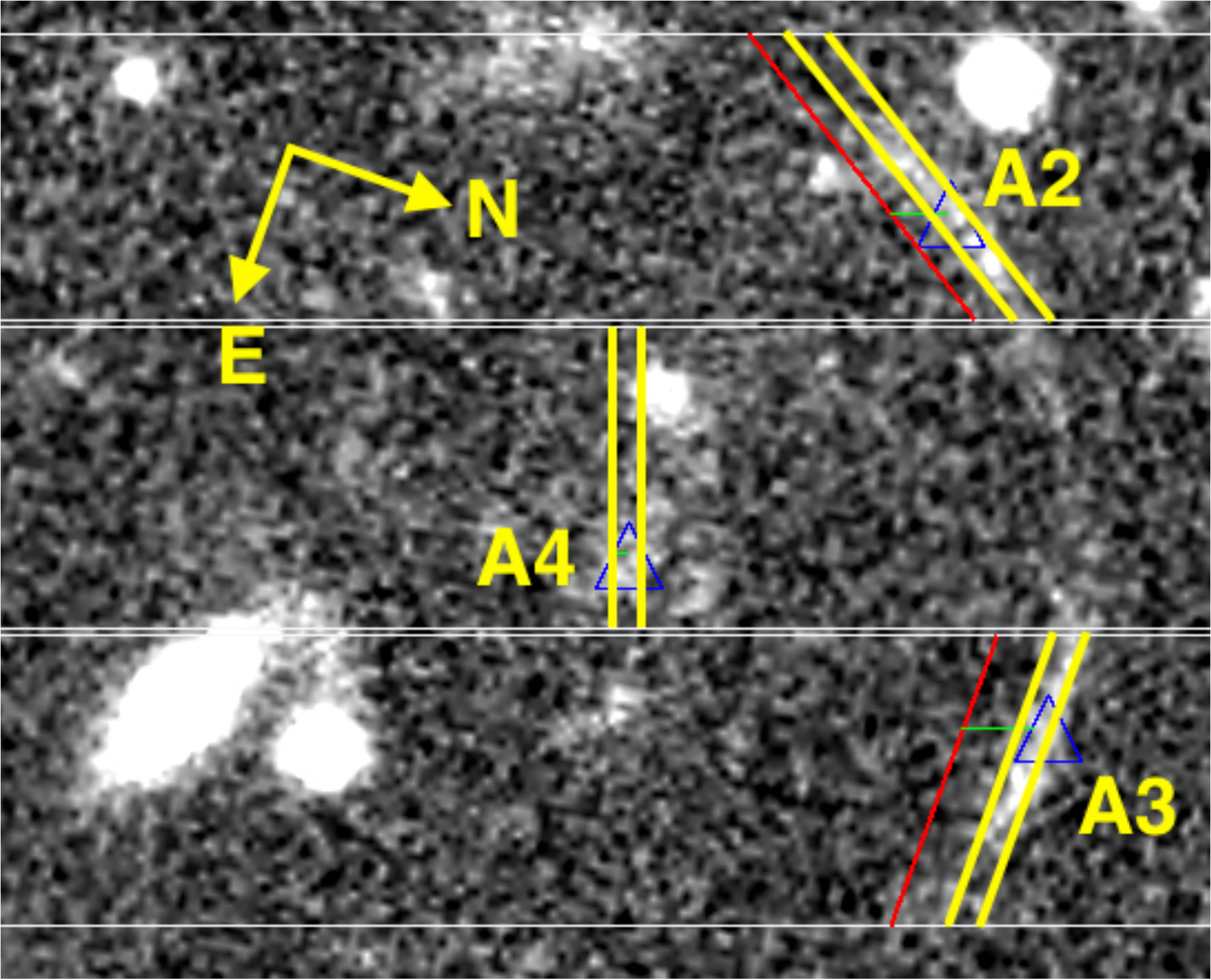}\\
    \includegraphics[width=0.6\columnwidth,clip=True]{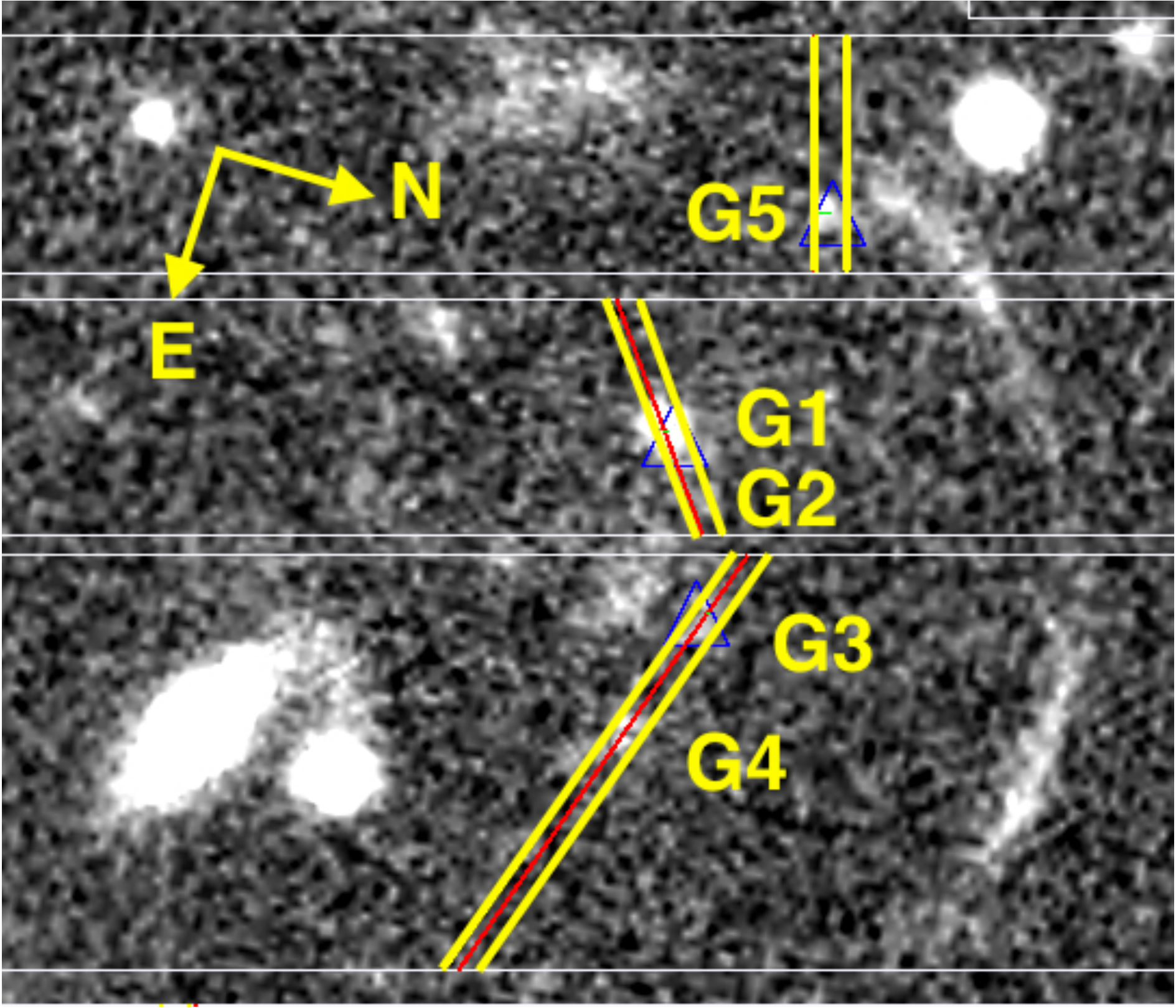}
    \caption {The Gemini GMOS masks for the spectroscopic observations. Top: Source mask. Bottom: Cluster mask.}
    \label{fig:masks}
\end{figure}

We targeted four of the red galaxies, labelled G1--G4 in Figure \ref{fig:colourimage}, in the center of the cluster including the Brightest Cluster Galaxy (BCG - labelled G1) along with another smaller galaxy close to arc A2 (labelled G5) and the arc features labelled A2--A4 using the multi-object mode on GMOS. We created two masks, one for the galaxies and the other for the arc features. \comment{In each case the mask was centered on the cluster and was rotated to a position angle of 80 deg West-of-North to maximize the slit placement.} The slits were $1\arcsec$ in width and of varying length in order to accommodate both the object and an amount of sky sufficient to perform reliable background subtraction. For some of the objects, we tilted the slits to maximize the captured flux.  The two masks are shown in Figure~\ref{fig:masks}. 

We observed the galaxies and arcs using two configurations with GMOS to cover the full wavelength range from $3250-10000 \angstrom$. We use the R150 grating in conjunction with the GG455 filter in order to obtain spectra with wavelength coverage $\sim4500-10000$\angstrom. If the source is a late-type emission line galaxy, this would allow us, in most cases, to detect [OII]3727 to $z\sim 1.7$, ${\rm H}_{\beta}$ to $z\sim 1.0$ and ${\rm Ly}_{\alpha}$ in the range $z\sim 2.7 - 7.2$. We use the B600 grating to obtain spectral coverage of $3250-6250$\angstrom, which would allow us to detect sources with $z > 2.0$ that emit ${\rm Ly}_\alpha$. The observations are summarized in Table~\ref{table:observationlog}. The seeing is taken from the Gemini data quality assessment.

An observing sequence consisted of a pair of exposures, followed by a flat-field taken with a quartz-halogen lamp and a calibration spectrum taken with a CuAr arc lamp. We then dithered to a different central wavelength to cover the gap between the CCDs and took a CuAr spectrum, followed by the flat-field exposure and then two more exposures. Exposure times were either 900 seconds or 840 seconds. Dividing the integration time into multiple exposures facilitates the removal of cosmic rays. The data were binned $2\times2$, giving effective dispersions of $0.1$ and $0.386$ nm/pixel for the B600 and R150 gratings, respectively. 

We used the Gemini IRAF package v2.16\footnote{\url{http://www.gemini.edu/sciops/data-and-results/processing-software}} to reduce all exposures.  In each system, for each wavelength dither, we first process the flat field using the {\tt gsflat} task (this includes subtraction of the bias). Each science exposure in a single dither is then reduced with {\tt gsreduce} (using the previously processed flat fields), and then the two exposures are combined with {\tt gemcombine}. Wavelength calibration and transformation are performed on each dither (using the {\tt gswavelength} and {\tt gstransform} tasks) before the pairs of dithers are coadded on a common wavelength scale to eliminate CCD chip gaps. We perform sky subtraction and 1D spectral extraction using {\tt gsextract}, which employs the {\tt apall} task. Feature identification and redshift estimation are performed using the {\tt rvsao} IRAF package \citep{kurtz98}.

{
\begin{table*}[!htb]
  \caption{Spectroscopic Observation Log  \label{table:observationlog}} 
  \centering 
  \footnotesize
  \begin{tabular}{ c c c c c c } 
    \hline\hline             
    Object &  UT Date       & Telescope-     & Grating  & Total Integration     & Seeing         \\ %[0.5ex] 
                      &                & Instrument     &               &     (hours)           &  ($\arcsec$)  
         \\ [0.5ex] 
\hline  \hline    
    Arcs       & 2014 Oct 20,24  & Gemini-GMOS    & B600 &      3                       & 0.86,0.66  \\
    Arcs       & 2015 Jul 17,19  & Gemini-GMOS    & R150 & 0.93                 & 0.9,1.06  \\
    Lens             & 2014 Oct  24   & Gemini-GMOS    &     R150 & 1                   & 0.66  \\
 [1ex]
 \hline \hline
\end{tabular}\\
\end{table*}
\subsubsection{Redshift of the lens}
The extracted un-fluxed 1-D spectra for the BCG (G1) is shown in Figure~\ref{fig:bcgspectrum}. The characteristic Ca H and K absorption lines present in early type galaxy spectra are visible in the spectra of all five galaxies at $\lambda=8192 \angstrom$ and $\lambda=8120 \AA$ respectively (indicated by the red lines in Figure~\ref{fig:bcgspectrum}). The redshifts obtained  for the five galaxies G1-G5 using a cross-correlation technique \citep[XCSAO]{kurtz98} are listed in Table~\ref{table:galaxyredshifts}. The BCG has a redshift of $z=1.0645\pm 0.0002$. The average redshift of the five galaxies is $z=1.0644$. In addition we obtained spectra for two additional galaxies that are likely to be members of the cluster as they have very similar redshifts. They are both about 1.5 arcmin from the cluster center with redshifts of $z = 1.0647$ and $z = 1.0620$. 

\subsubsection{Redshift of the source}
The extracted un-fluxed spectra and their $\pm 1 \sigma$ errors (green spectra) for the arcs A2 thru A4 are shown in Figure~\ref{fig:arcspectra} for the B600 grating. We detected continuum flux in all three arcs but no obvious emission line features. We do however observe a break in the spectrum at around $4100 \angstrom$ and absorption features at $4410 \angstrom$, $4517 \angstrom$ and $4720 \angstrom$. These features are indicative of those found in the spectra of a Lyman Break Galaxy (LBG) \citep{steidel} but with Ly$\alpha$ absorption rather than emission. We perform a cross-correlation \citep[XCSAO]{kurtz98} against the LBG template of \citet{shapley} for all three arcs. The results are listed in Table~\ref{table:arcredshifts}. The R value in Table~\ref{table:arcredshifts} provides a measure of the confidence of the redshift extraction and the uncertainty \citep{tonry}. The R values for A2 and A3 are both greater than 4 which is an indication of a secure redshift (as defined by \citet{kurtz98}). The R value for A4 is lower and reflects the lower signal to noise in the spectrum. However the presence of a break in the spectrum at around $4100 \angstrom$ consistent with the ones observed in A2 and A3 and the resulting redshift which is consistent with those obtained from A2 and A3 give us confidence that A4 is indeed a lensed image of the source galaxy. The mean redshift from the three arcs is $2.3880\pm0.0003$. Figure~\ref{fig:arcspectra} shows the locations of the expected spectral features for an LBG and the red spectrum is the LBG template of \citet{shapley} shifted by the measured redshift. The redshift of the lens and source imply a total mass within the Einstein radius of $\log_{10}(M/M_{\odot}) = 14.170 \pm 0.004$.

{
\begin{table}[!htb]
  \caption{Redshifts for A2-A4  \label{table:arcredshifts}}
  \centering
  \footnotesize
  \begin{tabular}{ c c c c c}
    \hline\hline
    Object &  RA       & DEC     & Redshift &  R-value\\
\hline  \hline
    A2      & 302.777796  & -52.468769   & $2.3875\pm 0.0002$ & 4.62\\
    A3      & 302.785149  &   -52.467079  & $2.3889\pm 0.0002$ & 5.34\\
    A4      & 302.783661  &  -52.471130    & $2.3875\pm 0.0004$ & 2.26\\
  \hline \hline
\end{tabular}\\
\end{table}

{
\begin{table}[!htb]
  \caption{Redshifts for G1-G5  \label{table:galaxyredshifts}} 
  \centering 
  \footnotesize
  \begin{tabular}{ c c c c } 
    \hline\hline             
    Object &  RA       & DEC     & Redshift \\
\hline  \hline    
    G1      & 302.78122  & -52.47105    & $1.0645\pm 0.0002$ \\
    G2      & 302.78244  & -52.47035    & $1.0737\pm 0.0002$ \\
    G3      & 302.78418  & -52.47032    & $1.0642\pm 0.0002$ \\
    G4      & 302.78605  & -52.47087    & $1.0514\pm 0.0002$ \\
    G5      & 302.77766  & -52.46994    & $1.0684\pm 0.0002$ \\
  \hline \hline
\end{tabular}\\
\end{table}

\comments{
\begin{table}[]
  \caption{DES photometry for arc A3  \label{table:photometry}} 
\centering 
\begin{tabular}{l c c c c c c} 
\hline  \hline   
Band  &$g$ & $r$ &$i$ &$z$ &$Y$ \\
\hline  \hline   
Magnitude& 21.88 & 21.49 & 21.61 & 21.75 & 21.84 \\
Error    & 0.041 & 0.056 & 0.085 & 0.138 & 0.628 \\
\hline  \hline   
\end{tabular}\\
\end{table}
}
\begin{figure}
  \centering
    \includegraphics[width=\columnwidth,clip=True]{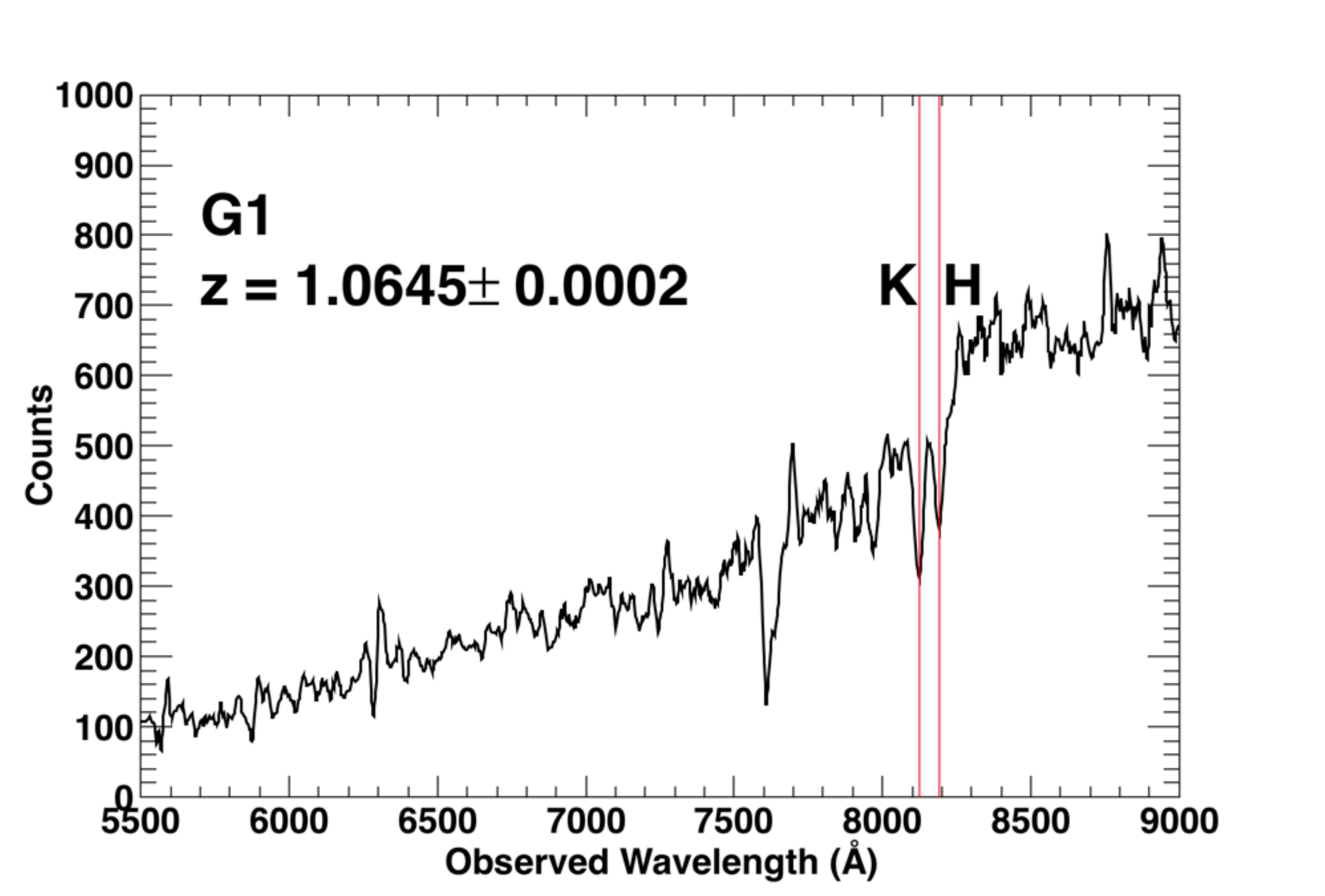}
    \caption {The extracted un-fluxed 1-D spectra for the BCG (G1). The Ca H and K absorption lines are indicated by the red lines.}
    \label{fig:bcgspectrum}
\end{figure}

\begin{figure*}
  \centering
    \includegraphics[width=\textwidth,clip=True]{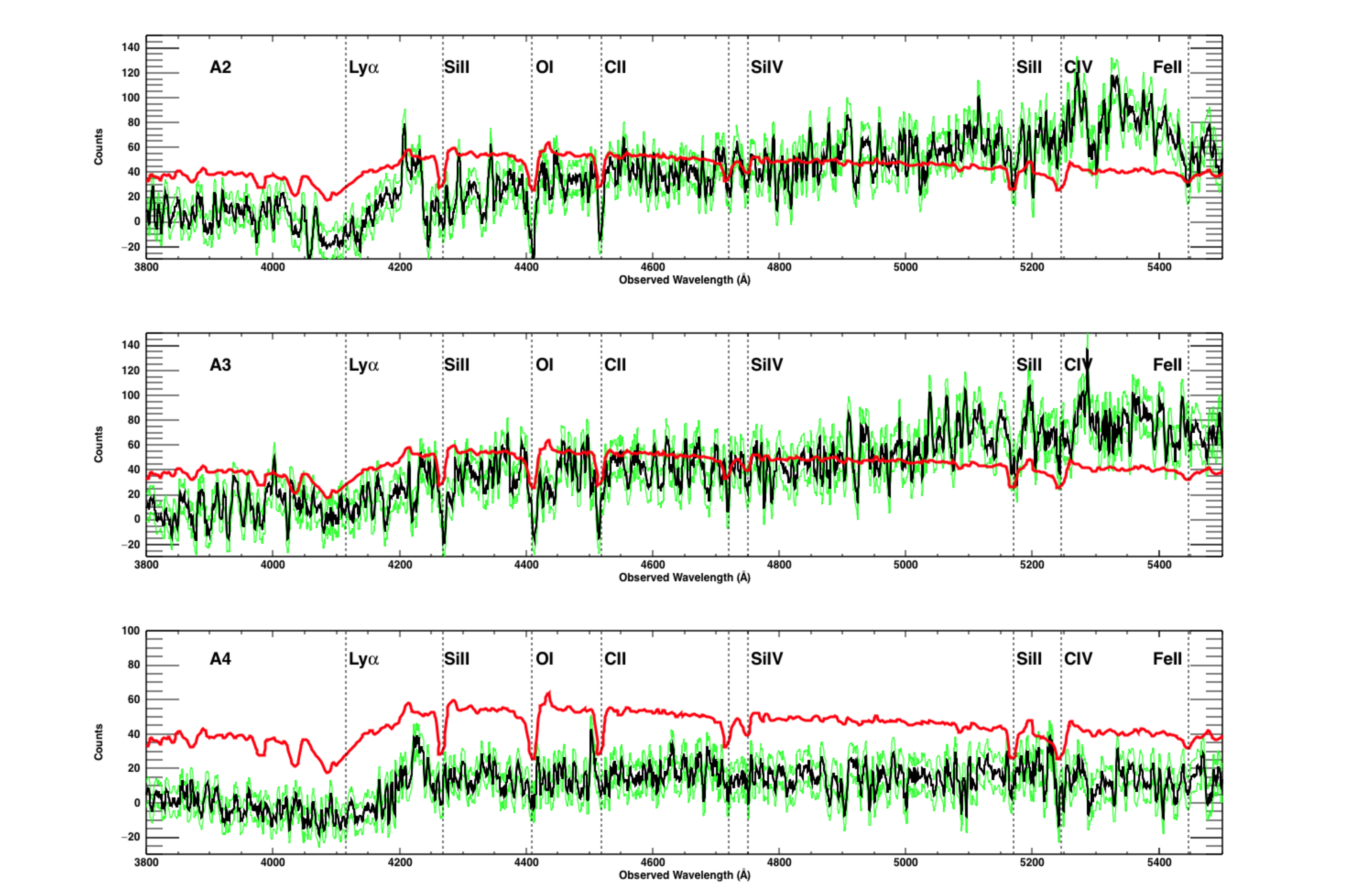}
    \caption {The extracted un-fluxed spectra for the arcs A2 thru A4. Absorption features from a Lyman Break Galaxy at $z=2.3875$ are indicated by the dotted lines.  The green spectra indicate the $\pm 1 \sigma$ errors from the spectral extraction. The red spectrum is the LBG template of \citet{shapley} shifted by the measured redshift.}
    \label{fig:arcspectra}
\end{figure*}

\section{Lens modelling of the lensed arcs and central image}
\label{sec:modelling}

We model J2011 with a cluster-scale dark matter halo plus subhalos to describe the baryons and dark matter associated with the cluster members and a single S\'ersic component light profile to describe the source. The presence of the central image gives us a unique opportunity to test the central density profile of the DM halo. We therefore fit the dark matter with a generalized NFW profile \citep{wyithe2001}. This allows us to test models with central cores or shallow cusps.}}

For the generalized NFW profile we assume the form
\be
\rho(r) = \frac{\rho_0}{r^{\alpha} (r_s^2+r^2)^{(3-\alpha)/2}},
\label{eq:dplprofile}
\ee
where $r$ is the three dimensional distance from the halo centre. The profile is characterised by an inner slope, $\alpha$, with the density falling as $r^{\alpha}$, an outer profile slope 3 and a break radius $r_s$. The case of $\alpha=1$ is approximately the NFW profile but with a slightly different behaviour in the turnover region. Using $(r_c^2+r^2)^{1/2}$ in the denominator rather than $(r_c+r)$ has the advantage that deflection angles for the spherical model can be calculated analytically \citep{munoz2001}. For computational efficiency we include ellipticity in the lensing potential rather than the density profile; this is a good approximation for almost spherical halos \citep{barkana}.

In addition to the cluster scale dark matter halo(es), mass is associated with the individual cluster members \citep{shaw2006}. We account for this mass by placing isothermal mass clumps in the locations where cluster members are observed. Using the z-band data, we fit the BCG light profile with an elliptical S\'ersic profile, and model the other cluster members with circular de Vaucoleurs profiles. For the BCG the ellipticity and position angle of the mass is fixed to those observed for the light profile. The other cluster members are modelled as isothermal spheres. In order to minimimize the number of free parameters in the model, we assume a constant mass-to-light ratio for all the cluster members. The Einstein radius of each component is proportional to the square root of the fitted z-band flux of each component: the constant of proportionality, $\tilde{\Upsilon}_{\mathrm{G}}$, is a free parameter of the model, $\tilde{\Upsilon}_{\mathrm{G}}$ is scaled in units such that it is also the Einstein radius of the BCG in arcseconds. The cluster members included in our model are shown in Figure \ref{fig:members}.

The final component of the lens model is an external shear, which allows for perturbative lensing from nearby perturbers \citep[e.g.][]{holder2003} and line-of-sight structures \citep[e.g.][]{collett2013,mccully2017}.

\begin{figure}
  \centering
    \includegraphics[width=\columnwidth,clip=True]{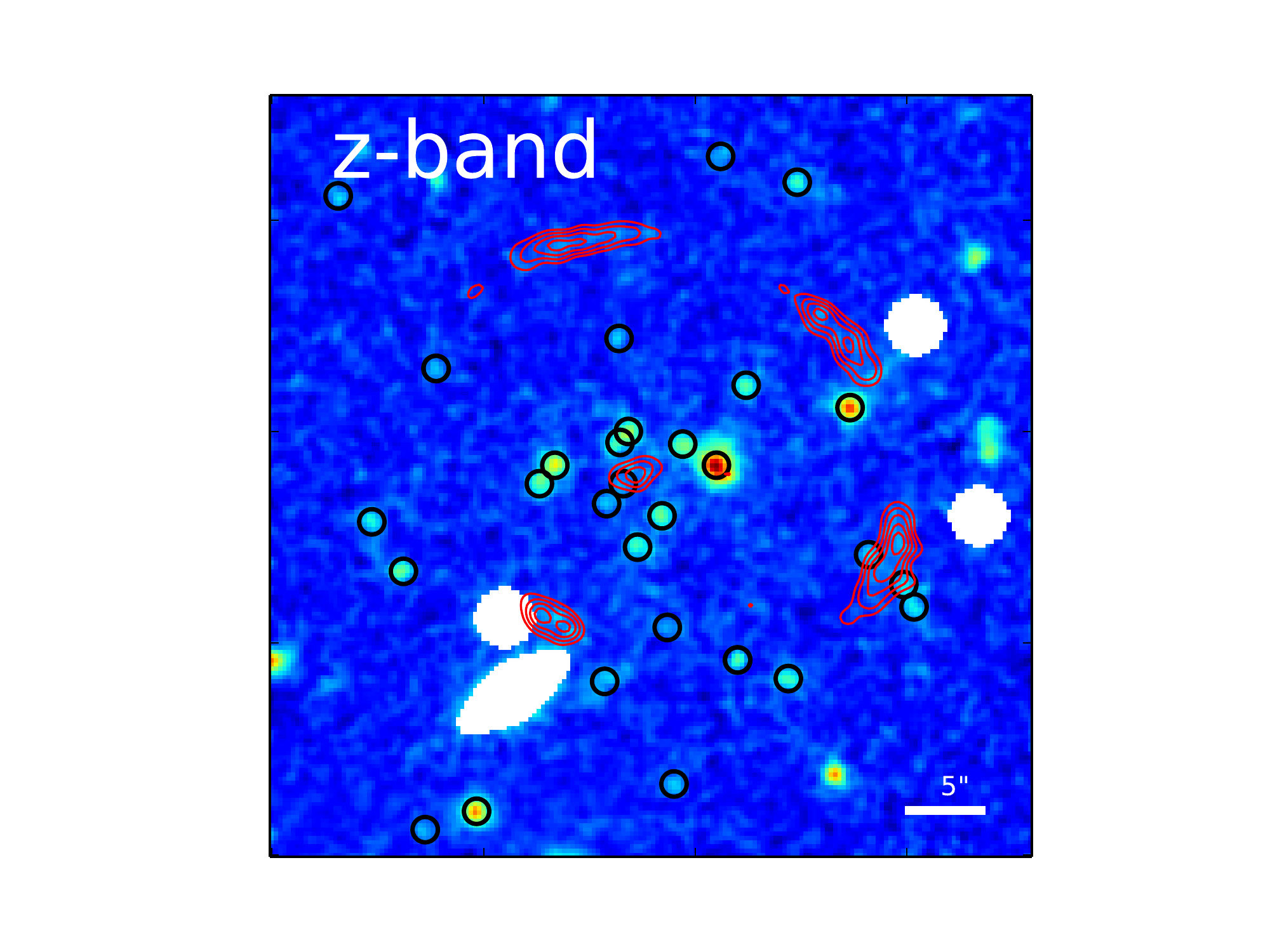}
    \caption {$z$-band image of J2011, with the arcs superimposed as red contours. The data has been filtered with a Gaussian filter (width 2 pixels) to pick out faint cluster members. Members included in the lens model are circled in black. The masked white regions are the locations of bright stars and a foreground galaxy.}
    \label{fig:members}
\end{figure}

\begin{figure*}
  \centering
    \includegraphics[width=\textwidth,clip=True]{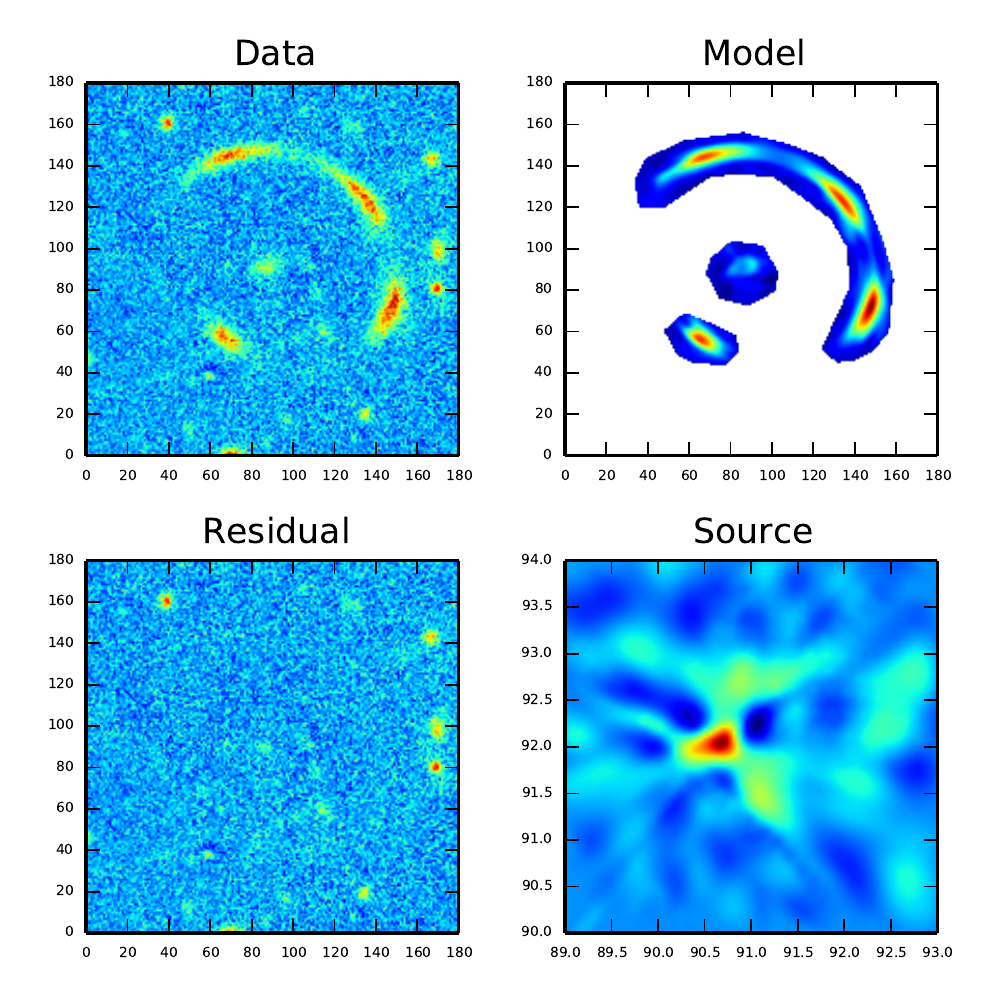}
    \caption {The best fit reconstruction of the g-band image of J2011, assuming a single gNFW halo for the dark matter. Clockwise from top left: 1) the foreground subtracted $g$-band data; 2) the best fit model of the lens; 3) the residual after subtracting the model from the data; 4) the best fit reconstruction of the unlensed source. \new{0,0 is at 20:11:10.611, -52:28:40.12 (J2000) with a pixel size of 0.263 arcseconds}}
    \label{fig:1gnfwfit}
\end{figure*}

\begin{figure*}
  \centering
    \includegraphics[width=\textwidth,clip=True]{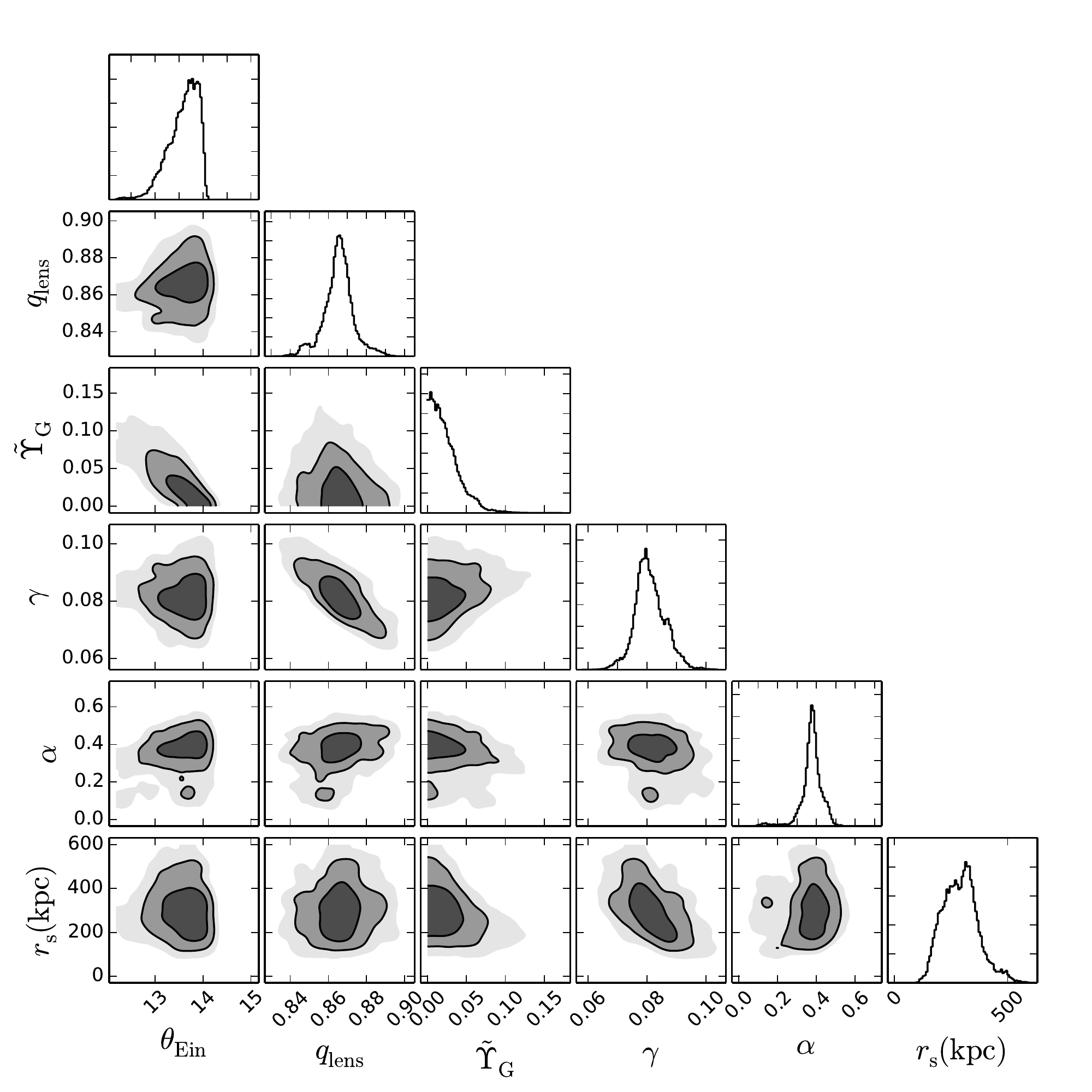}
    \caption {The marginalized 1 and 2 dimensional parameter constraints for the single dark matter halo model of J2011. The contours show the 68, 95 and 99 percent confidence regions respectively. $\theta_{\mathrm{E}}$ is the Einstein radius of the DM halo \footnote{Not the total Einstein radius of the arcs - since this is made up from the DM and the cluster members}. The flattening of the DM halo is $q_{\mathrm{lens}}$, $\alpha$ is the inner profile slope and $r_{\mathrm{s}}$ (kpc) is the scale radius of the gNFW halo. $\tilde{\Upsilon}_{\mathrm{G}}$ relates the observed $z$-band fluxes and the Einstein radiis of the cluster members, it is in units such that $\tilde{\Upsilon}_{\mathrm{G}}$ is the Einstein radius of the BCG in arcseconds.  The inner profile slope and the scale radius of the gNFW halo have only a mild covariance with the unshown parameters of the model.}
    \label{fig:corner}
\end{figure*}

\begin{figure*}
  \centering
    \includegraphics[width=\textwidth,clip=True]{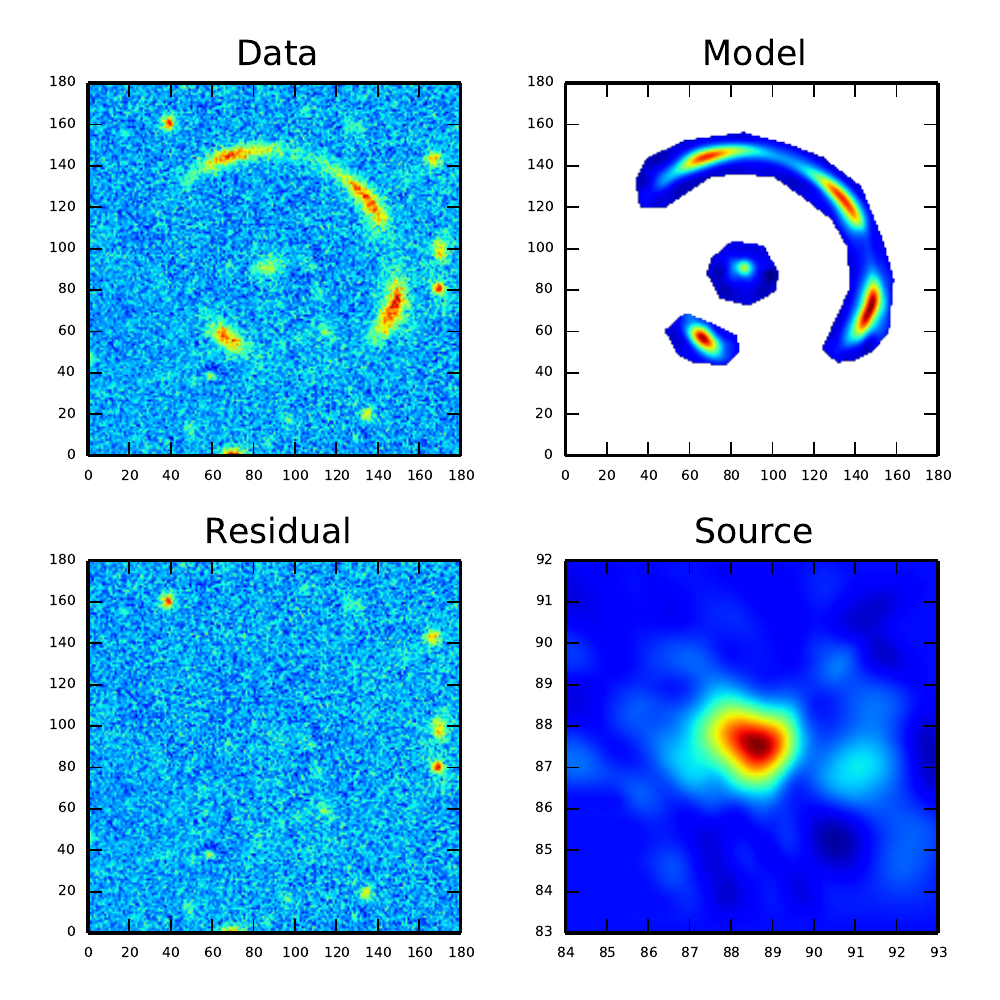}
    \caption {The best fit reconstruction of J2011. The same as Figure \ref{fig:2gnfwfit}, but assuming a 2 dark matter halo model for the lensing mass.}
    \label{fig:2gnfwfit}
\end{figure*}

\begin{figure*}
  \centering
    \includegraphics[width=\textwidth,clip=True]{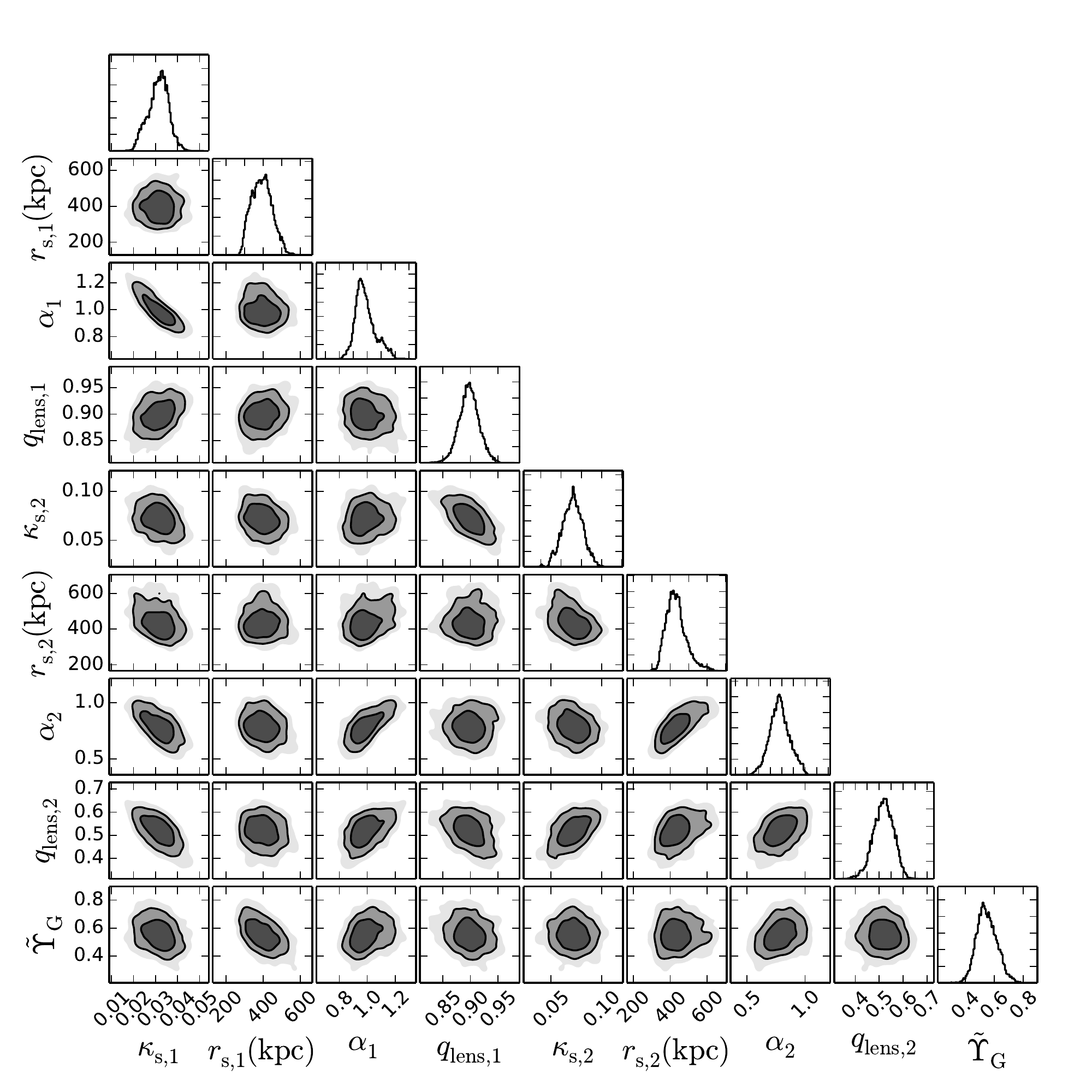}
    \caption {The marginalized 1 and 2 dimensional parameter constraints for the two dark matter halo model of J2011. The contours show the 68, 95 and 99 percent confidence regions respectively. Parameters with $_1$ refer to the primary clump and $_2$ refer to the secondary clump.}
    \label{fig:corner}
\end{figure*}

In order to fully exploit the information content of the central image it is necessary to attempt to reconstruct the light profiles of the lensed images. Full light profile fitting is more computationally challenging than conjugate point methods \citep[as used in][for example]{lenstool}, however it allows us to use the observed fluxes in many thousands of pixels as constraints on the lens model. 

Full light profile fitting requires some assumptions to be made about the unlensed source. Simply parameterized sources are quick to calculate and are robust against predicting images where none are observed, although they often require many non-linear source model parameters \citepeg{brewer2011}, and the extra sampling negates the computational benefits. More seriously the simplicity of the assumed light profile means that significant residuals are often present in regions of high magnification. These residuals cause simply parameterized sources to underestimate the uncertainties on the lens model (Collett \& Auger, In Prep). Pixellated sources are much more agnostic about the source profile, yielding better fits to the data and more robust estimates of the lens model parameters. However this is only true when the lens model is reasonably close to truth: The high computational cost of pixellated source modelling means that only a small amount of data can be reconstructed---the region including and immediately surrounding the lensed images---and lens models that predict images outside of the reconstructed region are not penalized.

We therefore first optimize our models towards an approximate best fit assuming the brightest observed pixel of the five lensed images are conjugate points, and use this lens model to initialize our full light profile fitting. We only model the parts of the image plane that contain flux from the source (shown in Figure \ref{fig:gsub}). To avoid the model predicting images outside of the masked region, we cast 100 pixels (outside of the mask) back onto the source plane and discard any model for which these pixels are within the source, defined as the smallest circle containing all five conjugate points. For the best fit model we also remodel the system with a much larger mask and verify visually that there are no extra images predicted.

 For the source profile we adopt a pixellated source model, following the semilinear approach of \citet{warren+dye}. We use an adaptive 50 by 50 grid of square pixels as detailed in \citet{collett2014} in order to avoid artificially breaking the mass-sheet degeneracy \citep{msd}.  We use curvature regularised sources as favoured by the analysis of \citet{suyu2006}; the regularisation encodes the prior that astrophysical sources are reasonably smooth. We follow the Bayesian prescription of \citet{suyu2006} and allow the data to tell us the optimal degree of source regularization for each iteration of the lens model.

Our lens model now has ten free non-linear parameters: two for the centroid of the halo ($x_{\mathrm{lens}}$,$y_{\mathrm{lens}}$), one for the Einstein radius of the main halo ($\theta_{\mathrm{E}}$), two for the ellipticity and position angle of the halo ($q_{\mathrm{lens}}$,$\theta_{q}$), the inner slope ($\alpha$) and break radius of the halo ($r_{\mathrm{s}}$), two parameters for the external shear ($\gamma$,$\theta_{\gamma}$), and one for the mass-to-light ratio of the cluster members ($\tilde{\Upsilon}_{\mathrm{G}}$). We probe the posterior of these non-linear parameters using a Markov Chain Monte Carlo method. We use the ensemble sampler of \citet{emcee}.

\subsection{ Results: A shallow cusp in the Dark Matter Halo}
\label{sec:1halo}

Applying the model of the previous section to J2011, we find that the DES data can be well reconstructed with an astrophysically reasonable source and a plausible mass distribution for the lens. The best fit reconstruction of the data is shown in Figure \ref{fig:1gnfwfit}. This model has an almost spherical dark matter halo with flattening $q=0.87$, a shallow inner density slope, $\rho \sim r^{-0.35}$ and a scale radius of 290 kpc, 2.2 percent of the mass within the Einstein radius is in subhalos. We see that the arcs and central image are reproduced with surprisingly small residuals given the simplicity of the mass model.  The model slightly under predicts the flux of the central image, indicating that the true central profile may be shallower than quoted. Due to the 1 arcsecond seeing of our imaging, the reconstructed source is not well resolved, however the model predicts that the source has one bright clump and possibly two extended nearby features; this irregularity is typical of the blue sources reconstructed in previous strong lensing studies \citepeg{brewer2011,shu2016}.

Our MCMC results show that the DM halo requires a shallow central density profile. Figure \ref{fig:corner} shows the marginalized two dimensional posterior for the DM halo properties. The inner slope, and scale radius are degenerate, however inner slopes steeper than $r^{-0.55}$ and scale radii smaller than 100 kpc are strongly excluded, indicating that the central profile deviates from the NFW prediction of $r^{-1}$ over a large radius. Marginalizing over the other parameters we derive 68 \% confidence intervals of $\alpha = 0.38 \pm 0.04$, and $r_s = 244^{+81}_{-64}$ kpc. The halo flattening is constrained to be $q_{\mathrm{lens}} = 0.877 ^{+0.006}_{-0.007}$. The external shear is $\gamma= 0.083 \pm 0.005$. And the constant of proportionality relating z-band flux to cluster member mass is $\tilde{\Upsilon}_{\mathrm{G}} = 0.02^{+0.02}_{-0.02}$, defined in units where $\tilde{\Upsilon}_{\mathrm{G}}$ is the Einstein radius of the BCG in arcseconds and the Einstein radii of the other members going as $\theta_i=\theta_{\mathrm{BCG}} (z^{\mathrm{Flux}}_i/z^{\mathrm{Flux}}_{\mathrm{BCG}})^{1/2}$.

\begin{table*}[]
  \caption{The infered parameters for the one halo model.  \label{table:results1}} 
\centering 
\begin{tabular}{l c c c c c c c c  c c c c c  c  c  c} 
\hline  
Parameter  &$x_{\mathrm{lens}}$&$y_{\mathrm{lens}}$& $\theta_{\mathrm{E}}$&$q_{\mathrm{lens}}$&$\theta_{q}$&$\alpha$&$r_{\mathrm{s}}$ (kpc) &$\gamma$&$\theta_{\gamma}$&$\tilde{\Upsilon}_{\mathrm{G}}$ \\
Inference &
$90.6^{+0.2}_{-0.3}$&
$91.6^{+0.2}_{-0.2}$&
$13.6^{+0.3}_{-0.4}$&
$0.866^{+0.006}_{-0.007}$&
$153^{+3}_{-3}$&
$0.38^{+0.04}_{-0.04}$&
$277^{+93}_{-74}$&
$0.081^{+0.006}_{-0.004}$&
$61.7^{+2.2}_{-1.6}$&
$0.02^{+0.02}_{-0.02}$\\
\hline  
\end{tabular}\\
\end{table*}

\section{An alternative model: Two merging Dark Matter halos}
\label{sec:2halo}

\begin{table*}[]
  \caption{The infered parameters for the two halo model.  \label{table:results1}} 
\centering 
\begin{tabular}{l c c c c c c c c c c c c c c c c c } 
\hline  
Parameter  &$x_{\mathrm{lens,1}}$&$y_{\mathrm{lens,1}}$& $\kappa_{s,1}$&$q_{\mathrm{lens,1}}$&$\theta_{q,1}$&$\alpha_1$&$r_{\mathrm{s,1}}$ (kpc)&$\tilde{\Upsilon}_{\mathrm{G}}$ \\
Inference & 
$82.9^{+0.5}_{-0.5}$&
$94.4^{+0.7}_{-0.7}$&
$0.032^{+0.004}_{-0.005}$&
$0.90^{+0.02}_{-0.02}$&
$-49.6^{+3.6}_{-4.3}$&
$0.98^{+0.08}_{-0.05}$&
$395^{+53}_{-54}$&
$0.55^{+0.08}_{-0.06}$
\\
\hline  
Parameter & $x_{\mathrm{lens,2}}$&$y_{\mathrm{lens,2}}$& $\kappa_{s,2}$&$q_{\mathrm{lens,2}}$&$\theta_{q,2}$&$\alpha_2$&$r_{\mathrm{s,2}}$ (kpc) \\
Inference &
$91.3^{+1.5}_{-1.6}$&
$73.3^{+2.1}_{-2.4}$&
$0.07^{+0.01}_{-0.01}$&
$0.52^{+0.04}_{-0.04}$&
$48.2^{+0.8}_{-0.9}$&
$0.79^{+0.09}_{-0.08}$&
$433^{+57}_{-46}$
\\
\hline  
\end{tabular}\\
\end{table*}

In the previous Section we found that the arcs and central image cannot be adequately reconstructed by a single (non-generalized) NFW halo, however the generalized NFW gets very close to reproducing the lens system, despite only having a small number of free parameters.
 
However the fact that gNFW model requires a much shallower profile than simulations predict and the slight underfitting of the central image implies that the true mass distribution may be more complicated than our simple gNFW model allows. One alternative hypothesis is that J2011 is a merger of two (or more) sub halos. We test this theory by adding a second gNFW clump to the model used in the previous section. \new{Since the halos are no longer guaranteed to be critical, we characterize them by their central convergence, $\kappa_{s,i}$, as defined in Equation 3 of \citet{munoz2001} and not by their Einstein radius.} We also do not include external shear in the two halo model as it is not required to reproduce the data. This model also prefers both halos to be approximately NFW. The model requires most of the mass to be in an almost spherical NFW halo, to the North West of the central image, with a less massive, highly flattened component to the South East. In this model twenty percent of the mass within the Einstein radius is associated with subhaloes. The central density of the two halos goes as radius to the powers $0.99^{+0.08}_{-0.06}$ and $0.79\pm{0.09}$ respectively. The break radii are $345^{+48}_{-42}$ kpc and $48.9^{+6.6}_{-5.6}$ kpc. We show the convergence map of the two dark matter halos in Figure \ref{fig:kappamap}.

\begin{figure}
  \centering
    \includegraphics[width=\columnwidth,clip=True]{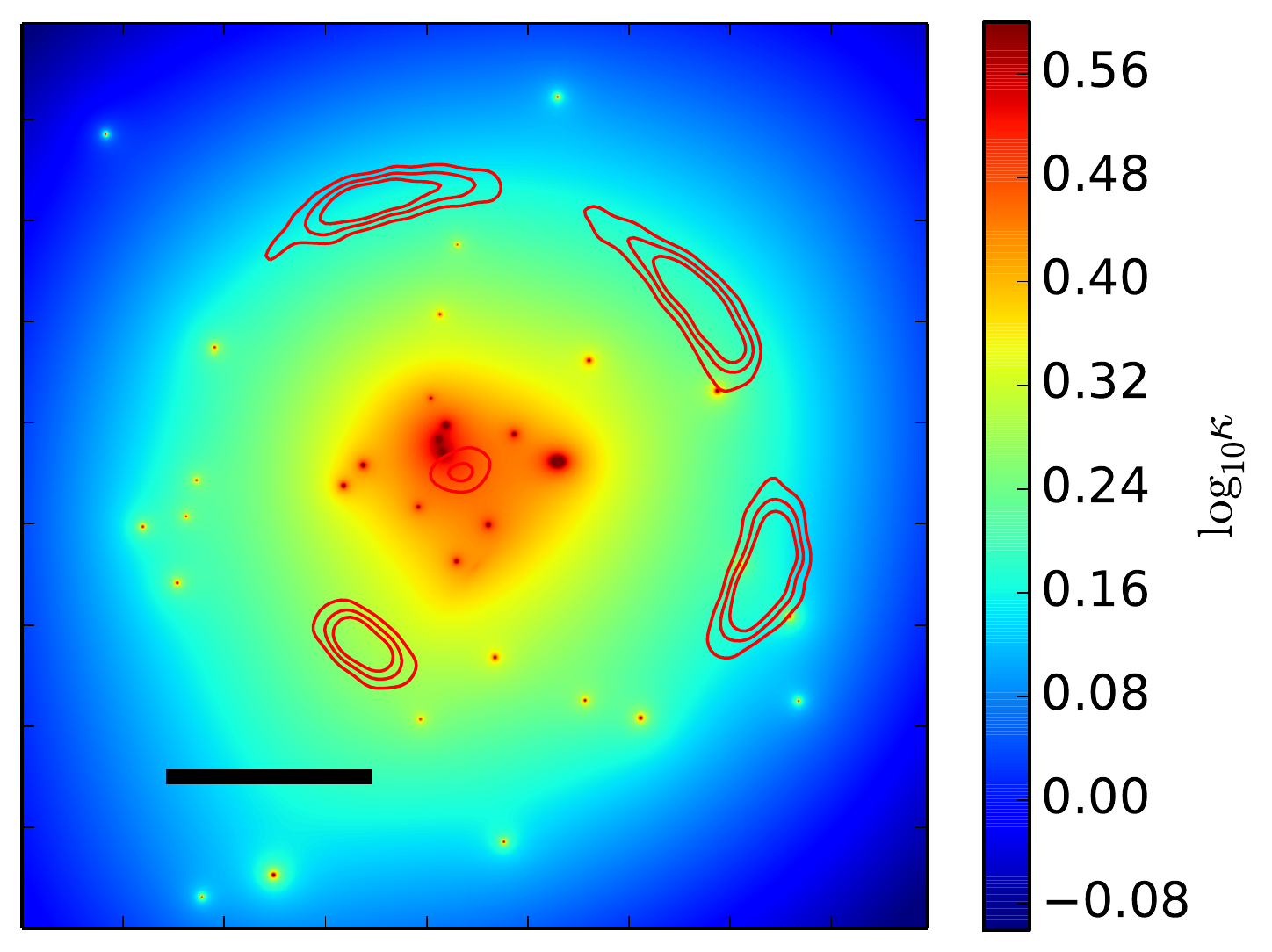}
    \caption {The mass distribution in J2011 as inferred from the two DM halo model. The red contours show the location of the lensed images. The black ellipses indicate the locations and flattenings of the two DM halos. The black bar indicates a 10 arcsecond scale.}
    \label{fig:kappamap}
\end{figure}

Figure \ref{fig:kappaslice} shows the convergence of the total density profile for various slices through the central image. Despite the fact that the DM halos and the cluster members are cuspy, the sum of the components gives a remarkably flat profile at the location of the central image.

\begin{figure}
  \centering       
    \includegraphics[width=\columnwidth,clip=True]{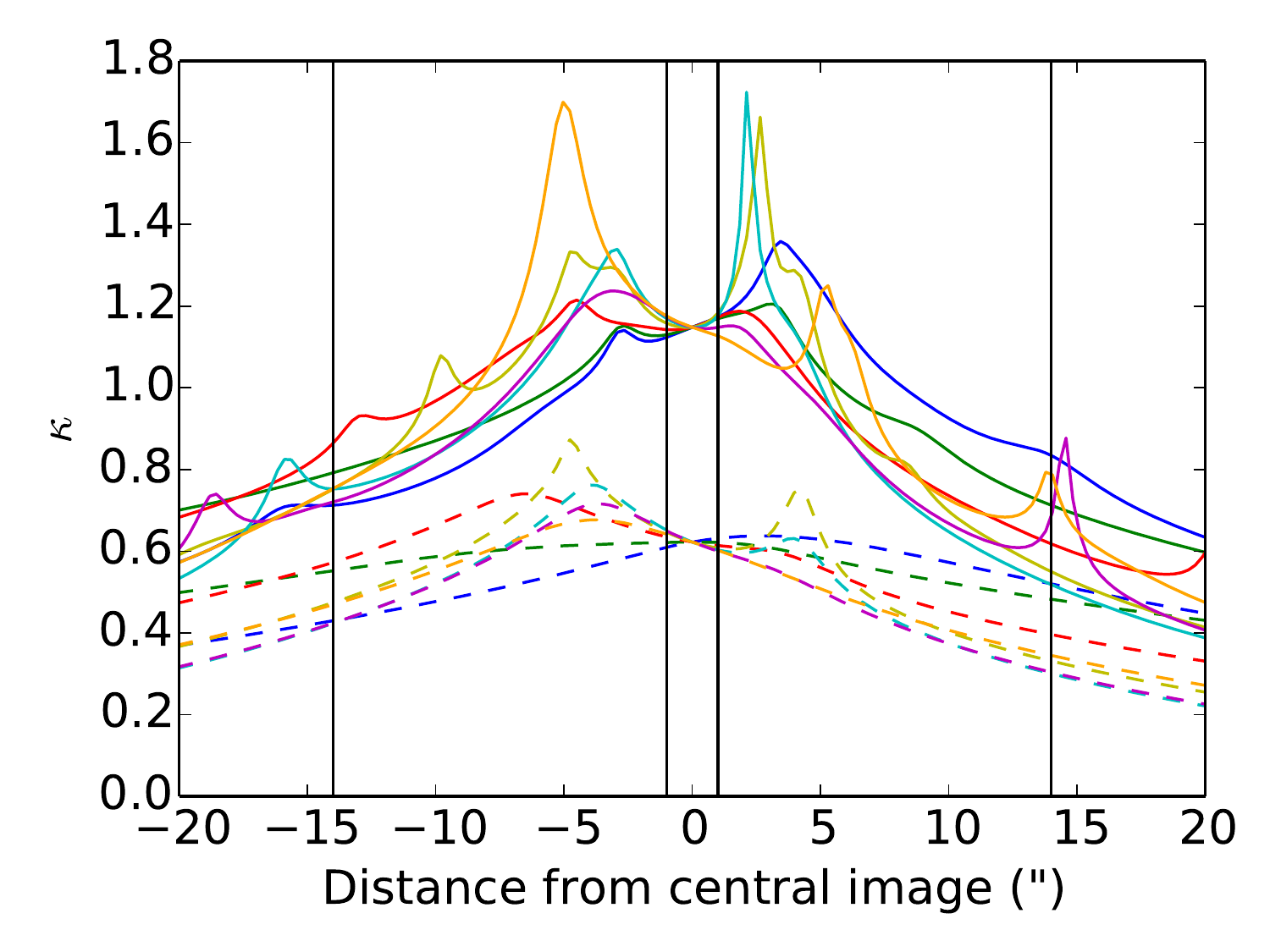}
    \caption {One dimensional slices through the surface mass distribution of J2011. Each slice goes through the central image, but the angles of the slices vary. The solid lines show the total surface mass distribution; the dashed lines show only the contribution from the inferred DM halos. The outer vertical lines indicate the Einstein radius. The inner vertical lines indicate the approximate size of the central image.}
    \label{fig:kappaslice}
\end{figure}

%,trim={15mm 5mm 10mm 10mm}

%%%%%%%%%%%%%%%%%%%%%%%%%%%%%%%%%%%%%%%%%%%%%%%%%%%%%%%%%%%%%%%%%%%%%%%%

\section{Interpretation of J2011 : a tension with $\Lambda$CDM or a merger of two NFW halos?}
\label{sec:interpretation}

\begin{figure}
  \centering       
    \includegraphics[width=\columnwidth,clip=True]{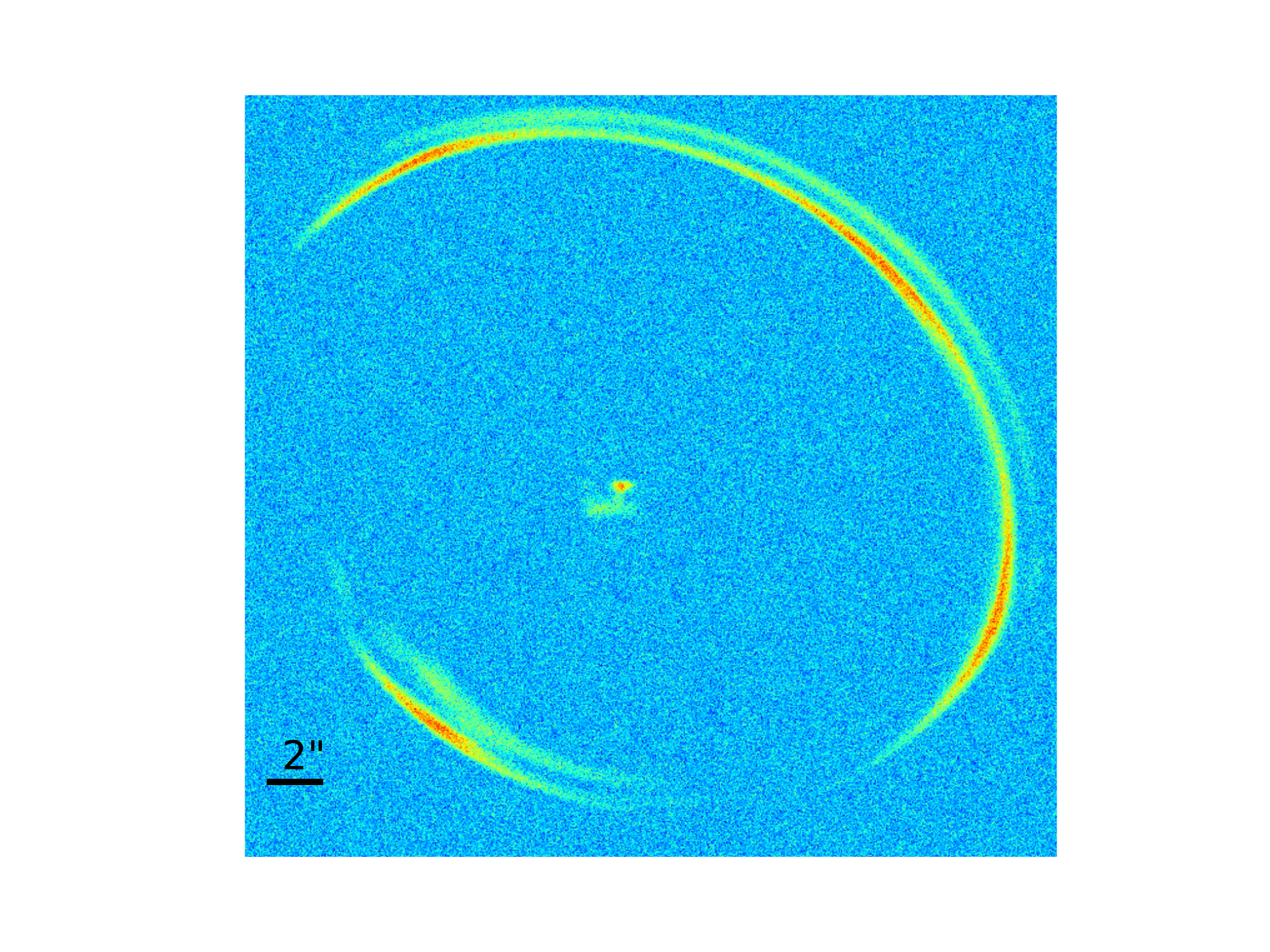}
    \includegraphics[width=\columnwidth,clip=True]{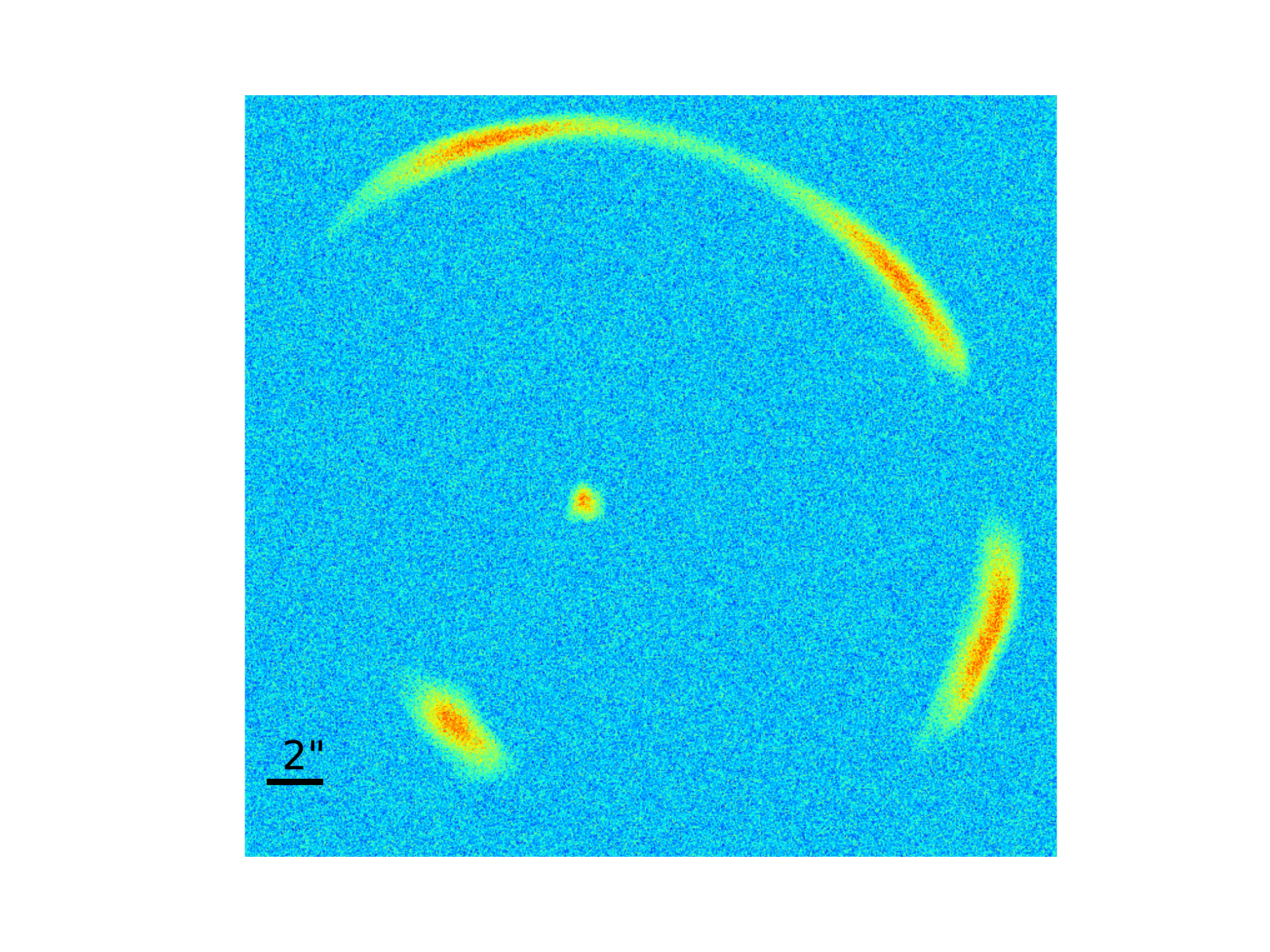}
    \caption {Simulated HST observations of J2011. The top image shows the expected HST image for the 1 dark matter halo model, the bottom image is for the two halo model. Noise levels are those expected for HST program GO 14630 (PI: Collett) scheduled for observations in cycle 24.}
    \label{fig:hst}
\end{figure}

Both the one and two halo models are able to reproduce the observed light profile of the arcs and the central image. The two halo model provides a somewhat better fit to the data. However, the log likelihood \new{(base 10)} of the residuals differ by only 17 totalled over 4892 pixels. The two halo model has 5 extra free parameters but also has a comparatively simpler source: the penalization term of the merit function \citep{suyu2006} prefers the two halo source by 19. Without the source penalization term, the Bayesian Information Criterion \citep{bic} would prefer the one halo model, however with the penalization term it prefers the two halo model.

The source penalization term of \citet{suyu2006} is however designed to prefer smoother sources, rather than precisely encode a known astrophysical prior about the clumpiness of sources, it is therefore unclear how to use the regularization term in quantitative Bayesian model selection. This same issue has been tackled in time delay cosmography \citep{wong2016}, with similarly unsatisfactory conclusions. We therefore conclude that with our data alone, both models are reasonable interpretations for this system.

Choosing between the two models thus reverts to astrophysical priors. From a physical perspective, neither model is totally satisfactory: in neither model is a DM halo centred on the BCG. For the one halo model the DM centroid is offset from the BCG by 30 kpc. For the two halo model, the main DM halo is centred around the clump of galaxies to the North West of the central image. The second DM halo is associated with the pair of galaxies to the South - although  given the extreme elongation of this halo it may be more likely that these galaxies are associated with two separate DM halos. The offset between the DM and the BCG in the one halo model is physically unrealistic in $\Lambda$CDM \citep{schaller2015}, however since $\Lambda$CDM should also not create giant dark matter cores, the two problems may be linked. In the two halo scenario it may not be surprising to see an offset between DM and the BCG as the system is not relaxed. \new{The small Einstein radii of the cluster members in the one halo model is also in tension with the typical Einstein radii of galaxy-scale lenses at lower redshifts. The one halo model implies a maximum velocity dispersion of only 60 km/s for the BCG whilst the two halo model gives a velocity dispersion of $219 \pm 15$ km/s. This may rule out the one halo model, but it could also be due to our simplistic assumptions about the cluster member density profiles.}

\new{Both our models assume a single lens plane; neglecting lensing by mass along the line of sight \citepeg{mccully2017}. There is a foreground spiral galaxy centred 5 arcseconds South East of Arc 5 which we have neglected throughout this work, although it may contribute to the external shear of the one halo model. There is no evidence that it significantly modifies the lensing potential at Arc 5, but even small changes in the geometry of a compound lens can have a significant effect in rare cases \citep{collettbacon2016}.}

 Conclusively discriminating between the two models presented in this work will require higher resolution imaging. HST program GO 14630 (PI: Collett) will observe J2011 in cycle 24 - and will trivially distinguish between the two models presented in this work\footnote{The HST data were taken shortly before submission of this paper. To avoid confirmation bias none of the authors of this paper (including the PI) had data access rights to view the HST data before this paper was submitted.}. We show simulated HST images for this program in Figure \ref{fig:hst}. The figures are generated assuming the sources are no more clumpy than those reconstructed from the low resolution DES data.

It is clear from Figure \ref{fig:hst} that whilst both models yield similar images in the DES data, the predictions for the HST data are very different, with both the shape of the central image and the radial width of the arcs providing tight constraints on the dark matter distribution. 

%%%%%%%%%%%%%%%%%%%%%%%%%%%%%%%%%%%%%%%%%%%%%%%%%%%%%%%%%%%%%%%%%%%%%%%%

\section{Conclusions}
\label{sec:conclusion}

In this work we have presented SPT-CLJ2011-5228, a giant system of arcs created by a cluster at $z=1$ possessing a unique central image. The cluster was already known from SPT \citet{song12}, but we have added followup in the form of DES imaging and Gemini spectroscopy.

With our data we have been able to spectroscopically confirm the cluster redshift to be $z_l=1.06$. The source is a Lyman Break Galaxy at $z=2.39$, implying a total mass within the arc of $\log_{10}(M/M_{\odot}) = 14.169 \pm0.004$, consistent with the SZ mass of $M_{500}=2.59 \pm 0.73 \times10^{14} h^{-1}_{70}\Msun$ \citep{bleem2015}

Irrespective of the source redshift we are able to model the density profile of the cluster. By describing the cluster as a single gNFW halo plus isothermal cluster members, we are able to reconstruct the arcs and central image so long as the DM halo is significantly shallower than NFW within the central $270^{+48}_{-76}$ kpc. We have also presented a two halo model that reproduces the images. In this model both halos are cuspy, but the total profile at the location of the central image is remarkably flat, allowing for the formation of a bright central image that is not demagnified.

At the signal-to-noise and resolution and of our imaging, the merger (two halo) model provides a somewhat better fit to the data than the shallow cusped (one halo) model, but at the expense of extra free parameters. Forthcoming HST imaging will be able to conclusively resolve between the shallow cusp and merger models. The stakes for these data are high, as the one halo model cannot be consistent with the concordance model of cold dark matter unless extreme amounts of baryonic feedback have re-sculpted the halo over hundreds of kiloparsecs. The forthcoming HST data will soon allow us to explain the unique central image of J2011, infer the dark matter profile in this system and perhaps shine a light on the nature of dark matter.

%%%%%%%%%%%%%%%%%%%%%%%%%%%%%%%%%%%%%%%%%%%%%%%%%%%%%%%%%%%%%%%%%%%%%%%%
%%  ACKNOWLEDGMENTS
%%%%%%%%%%%%%%%%%%%%%%%%%%%%%%%%%%%%%%%%%%%%%%%%%%%%%%%%%%%%%%%%%%%%%%%%

\section*{Acknowledgements}

We thank the referee for their helpful comments in improving upon the original manuscript.
We are grateful to Tomasso Treu and Wojciech Hellwing for fruitful discussions. TEC is grateful to Matt Auger for his significant contributions to the lens modelling code used in this work. We are grateful to Marusa Bradac and Matt Auger for encouraging us to investigate the merging halo scenario.

Funding for the DES Projects has been provided by the U.S. Department of Energy, the U.S. National Science Foundation, the Ministry of Science and Education of Spain, 
the Science and Technology Facilities Council of the United Kingdom, the Higher Education Funding Council for England, the National Center for Supercomputing 
Applications at the University of Illinois at Urbana-Champaign, the Kavli Institute of Cosmological Physics at the University of Chicago, 
the Center for Cosmology and Astro-Particle Physics at the Ohio State University,
the Mitchell Institute for Fundamental Physics and Astronomy at Texas A\&M University, Financiadora de Estudos e Projetos, 
Funda{\c c}{\~a}o Carlos Chagas Filho de Amparo {\`a} Pesquisa do Estado do Rio de Janeiro, Conselho Nacional de Desenvolvimento Cient{\'i}fico e Tecnol{\'o}gico and 
the Minist{\'e}rio da Ci{\^e}ncia, Tecnologia e Inova{\c c}{\~a}o, the Deutsche Forschungsgemeinschaft and the Collaborating Institutions in the Dark Energy Survey. 

The Collaborating Institutions are Argonne National Laboratory, the University of California at Santa Cruz, the University of Cambridge, Centro de Investigaciones Energ{\'e}ticas, 
Medioambientales y Tecnol{\'o}gicas-Madrid, the University of Chicago, University College London, the DES-Brazil Consortium, the University of Edinburgh, 
the Eidgen{\"o}ssische Technische Hochschule (ETH) Z{\"u}rich, 
Fermi National Accelerator Laboratory, the University of Illinois at Urbana-Champaign, the Institut de Ci{\`e}ncies de l'Espai (IEEC/CSIC), 
the Institut de F{\'i}sica d'Altes Energies, Lawrence Berkeley National Laboratory, the Ludwig-Maximilians Universit{\"a}t M{\"u}nchen and the associated Excellence Cluster Universe, 
the University of Michigan, the National Optical Astronomy Observatory, the University of Nottingham, The Ohio State University, the University of Pennsylvania, the University of Portsmouth, 
SLAC National Accelerator Laboratory, Stanford University, the University of Sussex, Texas A\&M University, and the OzDES Membership Consortium.

The DES data management system is supported by the National Science Foundation under Grant Number AST-1138766.
The DES participants from Spanish institutions are partially supported by MINECO under grants AYA2015-71825, ESP2015-88861, FPA2015-68048, SEV-2012-0234, SEV-2012-0249, and MDM-2015-0509, some of which include ERDF funds from the European Union. IFAE is partially funded by the CERCA program of the Generalitat de Catalunya.

Based on observations obtained at the Gemini Observatory (processed using the Gemini IRAF package), which is operated by the Association of Universities for Research in Astronomy, Inc., under a cooperative agreement with the NSF on behalf of the Gemini partnership: the National Science Foundation (United States), the National Research Council (Canada), CONICYT  (Chile), Ministerio de Ciencia, Tecnolog\'{i}a e Innovaci\'{o}n Productiva (Argentina), and Minist\'{e}rio da Ci\^{e}ncia, Tecnologia e Inova\c{c}\~{a}o (Brazil).

%%%%%%%%%%%%%%%%%%%%%%%%%%%%%%%%%%%%%%%%%%%%%%%%%%%%%%%%%%%%%%%%%%%%%%%%%%%%%%
%  APPENDICES
%%%%%%%%%%%%%%%%%%%%%%%%%%%%%%%%%%%%%%%%%%%%%%%%%%%%%%%%%%%%%%%%%%%%%%%%%%%%%%
\comments{
\appendix
}
%%%%%%%%%%%%%%%%%%%%%%%%%%%%%%%%%%%%%%%%%%%%%%%%%%%%%%%%%%%%%%%%%%%%%%%%%%%%%%
%  REFERENCES
%%%%%%%%%%%%%%%%%%%%%%%%%%%%%%%%%%%%%%%%%%%%%%%%%%%%%%%%%%%%%%%%%%%%%%%%%%%%%%
%\twocolumn
% MNRAS does not use bibtex, input .bbl file instead. 
% Generate this in the makefile using bubble script in scriptutils:

% bubble -f paper-lcr.tex references.bib 
% \input{paper-lcr.bbl}

% \bibliographystyle{apj}
% \bibliography{references}

%%%%%%%%%%%%%%%%%%%%%%%%%%%%%%%%%%%%%%%%%%%%%%%%%%%%%%%%%%%%%%%%%%%%%%%%%%%%%%

\label{lastpage}
%\bsp

\end{document}